# Shearing Mechanisms of Co-Precipitates in IN718

Christopher H Zenk[a,*,1], Longsheng Feng[a], Donald McAllister[a], Yunzhi Wang[a], Michael J Mills[a,*]

[a]Department of Materials Science and Engineering, The Ohio State University, 2041 College Road, Columbus, OH 43210, USA

[*]Corresponding author

**Abstract**

The Ni-base superalloy 718 is the most widely used material for turbomachinery in the aerospace industry and land-based turbines. Although the relationship between processing and the resulting properties is well known, an understanding of the specific deformation mechanisms activated across its application temperature range is required to create more mechanistically accurate property models. Direct atomic-scale imaging observations with high angle annular dark-field scanning transmission electron microscopy, complemented by phase-field modeling informed by generalized stacking fault surface calculations using density functional theory, were employed to understand the shear process of γ″ and γ′/γ″ co-precipitates after 1 % macroscopic strain at lower temperature (ambient and 427 °C). Experimentally, intrinsic stacking faults were observed in the γ″, whereas the γ′ was found to exhibit anti-phase boundaries or superlattice intrinsic stacking faults. Additionally, the atomically flat γ′/γ″ interfaces in the co-precipitates were found to exhibit offsets after shearing, which can be used as tracers for the deformation events. Phase-field modeling shows that the developing fault-structure is dependent on the direction of the Burgers vector of the $\frac{a}{2}\langle 110 \rangle$ matrix dislocation (or dislocation group) due to the lower crystal symmetry of the γ″ phase. The interplay between γ′ and γ″ phases results in unique deformation pathways of the co-precipitate and increases the shear resistance. Consistent with the experimental observations, the simulation results indicate that complex shearing mechanisms are active in the low-temperature deformation regime and that multiple $\frac{a}{2}\langle 110 \rangle$ dislocations of non-parallel Burgers vectors may be active on the same slip plane.

## 1 Introduction

Ni-base superalloy 718 has been prominently used in aerospace applications since the 1960s [1]. It has been favored as a structural material for its yield strength retention at elevated temperatures as well as its weldability [2]. Due to its extensive use over many decades, the relationship between processing parameters and the resulting properties has become well understood. However, still poorly understood is the interplay between processing and the resulting microstructure, and how that microstructure (in particular, the presence of co-precipitates) dictates the deformation mechanisms operative over a range of temperatures relevant for industrial applications [3]. The characterization and phase-field modeling work presented here were produced in an effort to understand the complex deformation mechanisms present in 718. The results are being used to inform the creation of a mechanistically accurate yield strength model and will

---

[1] Present address: Department of Materials Science and Engineering, Institute WTM, Friedrich-Alexander-University Erlangen-Nürnberg, Martensstr. 5, D-91058 Erlangen, GERMANY

presently focus on lower temperature deformation.

Alloy 718 is primarily strengthened by the γ″ phase [4]. The γ″ phase has an ordered tetragonal (D0$_{22}$) crystal structure[2] and is based on the chemical stoichiometry of Ni$_3$Nb [5]. The γ′ phase (Ni$_3$Al, L1$_2$[2], cubic) contributes to the strengthening of alloy 718 to a lesser degree. Both phases are coherently embedded in a disordered face-centered cubic (fcc) γ matrix. The γ″ phase forms as lenticular particles of three different variants in which the [001]$_{γ″}$ for each variant is parallel to each of the principle <100> directions[3] in the γ matrix and γ′ precipitates [4,6]. Composite particles comprised of both γ′ and γ″ phases are also observed [7].

Researchers have strived to address the deformation mechanisms present in alloy 718. Early attempts at predicting deformation pathways within the γ″ phase suggested that quadruplets of like-signed $\frac{a}{2}\langle 110 \rangle$ matrix dislocations would be necessary to return the γ″ back to its original crystal symmetry in all three variants [4]. More contemporary investigations have shown the presence of stacking faults isolated in the γ″ phase as well, sometimes extending into the matrix following some heat treatments [8,9]. Localized microtwinning is also prevalent in some overaged samples [3,8]. It is important to note that much of the previous work has not been performed on peak-aged samples [4,8]. Due to the diversity of deformation mechanisms reported previously in the literature, there is a need to develop a cohesive view of deformation mechanisms present in commercially heat-treated material across its usable temperature range.

A generalized stacking fault (GSF) energy surface has been calculated for the γ″ phase previously using density functional theory (DFT) [10]. The GSF energy surface predicts the temperature-invariant (0K) fault energy for a single{ 111 } shear plane. It has been shown that the GSF energy surface for the γ″ phase has lower symmetry than that for the γ′ phase [10], so a single $\frac{a}{2}\langle 110 \rangle$ matrix dislocation would shear each variant along a unique shear path. Additionally, the formation of unstable complex stacking faults (CSF) and anti-phase boundary-like (APB-like) faults, instead of stable fault structures in the γ′ phase, leads to the possibility of complex shearing mechanisms that have not previously been recognized or explored [11]. Preliminary work using diffraction contrast scanning transmission electron microscopy (DF-STEM) imaging has indicated that single $\frac{a}{2}\langle 110 \rangle$ shearing/looping is generally not operative following room temperature deformation in a commercially heat-treated condition [11]. Therefore, this investigation focuses on the use of HAADF STEM imaging to examine the atomic configurations of sheared particles in order to validate the specific shear paths predicted through phase field modeling.

## 2 Experimental Methods

Bulk alloy 718 samples were received from Honeywell Aerospace. They had been solutionized at 968 °C and subsequently precipitation-hardened by the traditional dual step commercial heat treatment of an 8 h hold at 718 °C followed by a cooling step at -60 °C/h to the final hold at 620 °C for 8 h. The final furnace-cool occurred over the course of 1 h. All experiments have been conducted in this condition, unless specified otherwise. To obtain coarser particles for EDS analysis, a sample was also overaged for 21 h at 750 °C.

---

[2] D0$_{22}$ and L1$_2$ denote crystal structures in *Strukturbericht* notation.

[3] To avoid confusion, we refer to crystallographic directions and planes using the indexing of the cubic γ and γ′ phases, even when discussing dislocations in the γ″ phase (instead of the tetragonal indexing its crystallography would require).

Multiple tensile specimens were extracted from various positions in a forging. Tensile samples were deformed at 22 and 427 °C at a strain rate of $5 \times 10^{-3}$ /s to a total of 1 % plastic strain.

TEM foils were extracted from the gauge length by using a high-speed precision cut-off machine. The samples were then polished to 300 µm in thickness using 240 grit polishing paper followed by incremental polishing steps to a thickness of 130 µm with 800 grit paper. Subsequently, 3 mm disks were extracted using a coring slurry drill with 320 grit boron carbide powder. The final disks were polished to 90 µm in thickness using 1200 grit polishing paper. The disks were then electropolished to electron transparency using a *Struers* dual-jet electropolishing unit with a solution of ~20 % perchloric acid and ~80 % methanol cooled to -40 °C with an applied potential of 10.5 V.

The complexity of diffraction contrast imaging of dislocation groups, as well as the obscuring effect that the precipitate strain fields have on diffraction contrast images, are significant challenges for conventional TEM approaches and characterizing deformation mechanisms [11]. In this case, atomic-scale imaging using high angle annular dark-field scanning transmission electron microscopy (HAADF STEM) can provide a more direct measurement of shear events and the interaction of slip with monolithic and composite particles. In this work, HAADF STEM was employed on an *FEI Probe Corrected Titan3™ 80-300 S/TEM* at a camera length of 73 mm and below to study the size and morphology of particles. The intensity of the HAADF STEM signal is sensitive to the atomic number "$Z$", so Nb-rich γ″ particles are identifiable due to their higher intensity. The probe corrector compensates for high order aberrations in the electron beam and provides sub-angstrom resolution so that individual atomic columns can be clearly resolved along several zone axes.

Energy dispersive spectroscopy (EDS) was utilized to investigate the complex morphology of the γ′/γ″ co-precipitates. It was performed on an *FEI Image Corrected Titan3$^{TM}$ G2 60-300 S/TEM* at a camera length of 73 mm or less. The EDS mappings were collected using an *FEI ChemiSTEM Super-X* detector with a beam optimized to give 5-15 kcps to give atomic resolution mapping. Additional EDS measurements to resolve potential elemental segregation to the γ′/γ″ interface were performed using *a Thermo Fisher Scientific$^{TM}$ Themis* STEM operating at 120 kV accelerating voltage.

## 3 Microstructure and Fault Characterization

### 3.1 Precipitate Microstructures

The microstructure of the materials examined in this study consists of an average grain size of around 100 µm, and a significant volume fraction (~3.5 %) of δ phase along grain boundaries. The focus of the present study is to elucidate the manner by which shearing of the important intragranular strengthening phases occurs at low and intermediate temperatures. A combination of higher magnification EDS

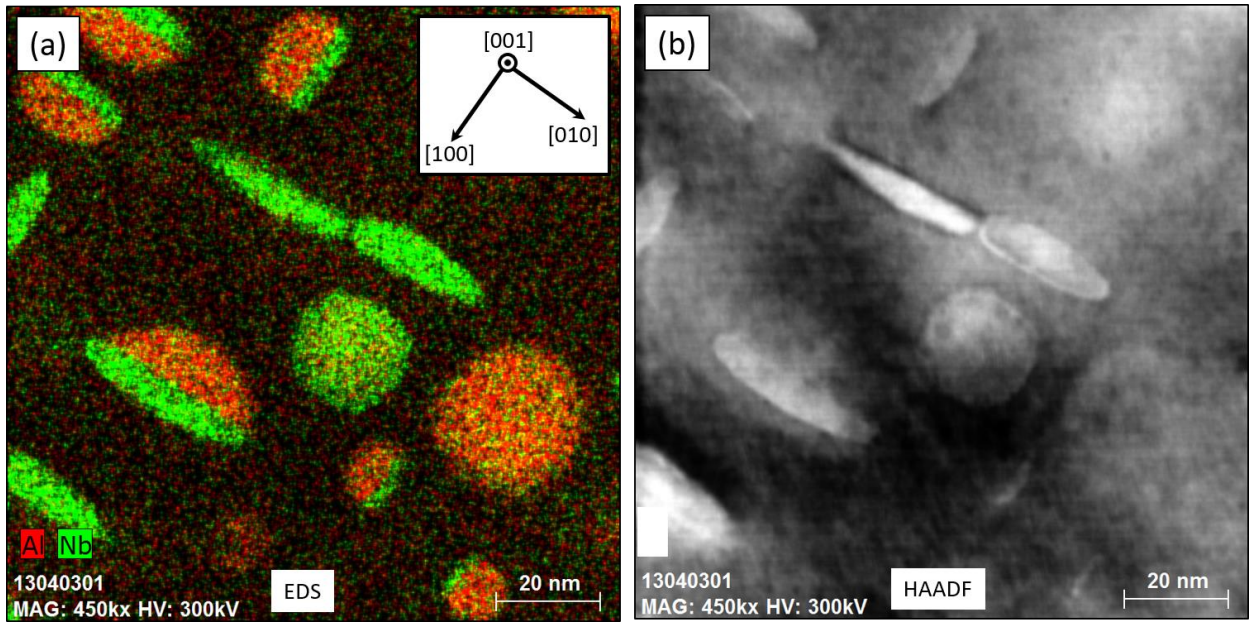

*Figure 1: (a) An EDS map of an overaged sample (750 °C / 24 h) obtained along [001]g showing the presence of all three variants of the γ″ precipitates demonstrating that the disk face normal being parallel to the three principle <100> directions of the γ matrix, and the accompanying HAADF STEM scan (b). Several examples of composite γ′/ γ″ particle are also observed with the interface planes perpendicular to [100] and [010].*

mapping and HAADF imaging of the typical precipitate structures are shown in Figure 1.

Both Al-rich γ′ (red) and Nb-rich γ″ phases (green) are observed as monolithic and composite particle configurations in the EDS map of Figure 1a, consistent with earlier analysis by Phillips, et al [7]. The presence of all three possible orientation variants of the disk-shaped γ″ precipitates can be surmised based on their elongation along [100] and [010] directions. The particles are expected to elongate normal to the c-axis of the tetragonal unit cell [12]. The Nb-rich features with squircular appearance are presumed to be γ″ precipitates viewed parallel to their c-axis. Composite particles are also common for this heat treatment and are readily identified in the EDS maps. When maps are obtained along a <001>$_\gamma$ orientation, then two variants of the composite particles can be clearly distinguished since the preferred {010} interface between γ′ and γ″ is then viewed "edge-on" and produces a distinct compositional interface.

The atomic ordering contrast in HAADF images of the γ″ particles has proven useful in tracking the cumulative shear history of a given slip plane, and has been utilized extensively in this study. Strong superlattice contrast is obtained for <001> zone axis images of the γ″ phase, as illustrated in Figure 2 for a [100]$_{\gamma″}$ zone axis.

Since Nb atoms sit at the corners and center of the $D0_{22}$ unit cell, the average intensity of Nb-rich atomic columns in the $(001)_{\gamma″}$ planes is easily distinguishable from the neighboring Ni-rich columns when the viewing direction is parallel to a {100}. However, when e.g. viewed along a ⟨110⟩ zone in HAADF STEM mode, the average atomic number of individual atomic columns is similar and obscures superlattice contrast. Imaging the defect structures to be discussed below along ⟨110⟩ zone axes is particularly important since shearing events occurring on two different {111} type planes can be viewed with these shearing planes "edge-on," thus clarifying the analysis.

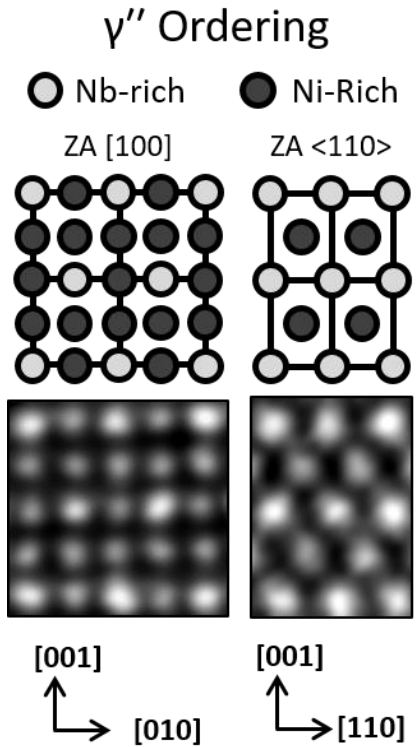

Figure 2: Schematic representations and corresponding HAADF-STEM micrographs of the γ'' phase when viewed along the [001] and ⟨110⟩ zone axes.

A distinct feature associated with the co-precipitated composite γ'/γ'' particles has been essential for the analysis. This feature is the atomically flat plane with low intensity in the HAADF signal that exists at the interface between γ' and γ'', as can be seen in Figure 3.

This feature will subsequently be referred to as low-intensity plane, or LI plane for short. The intensity of this plane is low, even relative to the adjacent γ' phase, and should be associated with an occupancy by solutes of a lower average atomic number in this plane. In order to clearly correlate the LI plane with compositional fluctuations at the interface, an atomic resolution EDS scan was performed on the γ''/γ' composite particle. Net EDS intensity for Al and Nb signals are shown in Figure 3c and d. An intensity profile was integrated normal to the LI plane to show the change in the HAADF, Nb, and Al signals, Figure 3e. The LI plane clearly corresponds to a local excess of the local atomic number element Al (ca. 20 % higher EDS intensity), explaining the lower intensity of the interfacial plane in the composite. Earlier atom probe investigations by Geng et al. are in agreement with our findings and also provide evidence of a slight excess of Al at the phase boundary [13].

It should also be noted that the γ' phase is difficult to distinguish from the γ matrix as the net atomic numbers of both phases are similar. Nevertheless, the presence of γ' can be directly inferred from the existence of the atomically-flat [001] interface between γ' and γ'' phases.

As will be discussed in subsequent sections, the LI plane only deviates from an atomically flat configuration when the co-precipitate has been sheared by a deformation event, or when there exists a step (growth ledge) in the interface plane associated with a displacement of the interface during particle growth. As exemplified in the HAADF image of Figure 4, a growth ledge can be readily differentiated from shear ledges because – contrary to the shear that is caused by typical dislocations in this material – the former have heights that are equivalent to a full γ'' unit cell, and they do not extend to both interfaces of the shown dual-lobed composite particle.

The LI plane is apparent in HAADF mode when viewing a γ'/γ'' composite particle interface edge-on, and can be seen in both ⟨100⟩ and ⟨110⟩ matrix orientations. It consistently exists in all composite particles and is a reliable reference for the analysis of local shear events: the displacement of the LI plane can be used as an atomic-scale displacement tracer for the total in-plane shear from dislocations that have passed on a {111} slip plane during deformation. The inherent limitation of this technique is the fact that the STEM micrographs provide a two-dimensional projection of the three-dimensional volume, so the out-of-plane shear cannot be determined from a single zone axis image.

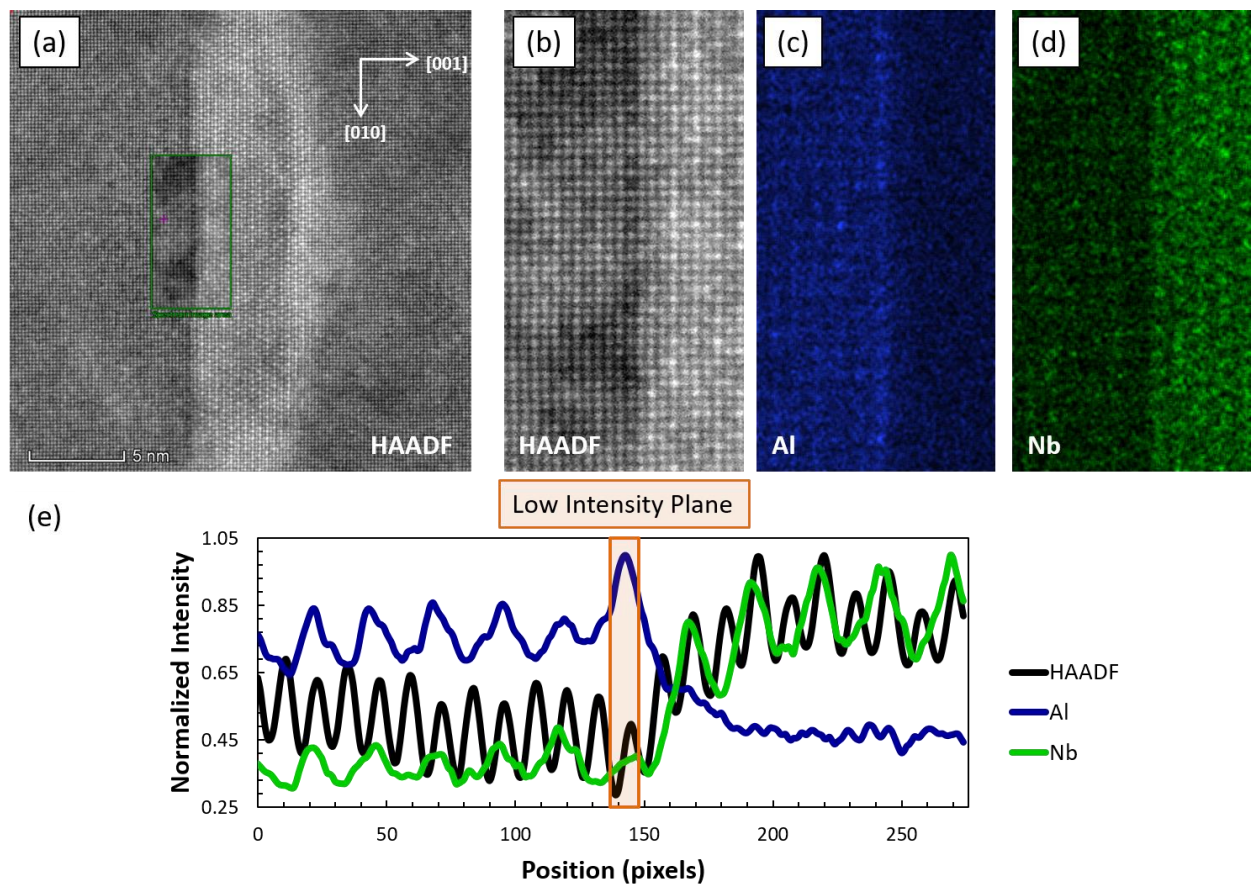

*Figure 3: High-resolution HAADF images (a and b) of a composite particle viewed along a [100] zone axis. An atomic-resolution EDS analysis of the low-intensity plane (c and d) and the corresponding integrated intensity profile (e) show an increased Al-concentration at the γ″/γ′ interface, coinciding with the low intensity observed in the HAADF micrograph.*

### 3.2 Stacking Faults in the γ″ Phase

Before presenting examples of measured shear events using HAADF imaging, it is useful to consider the variety of possible stacking faults that can occur in the γ″ phase since these are inherently more complex than those in the higher symmetry and more familiar γ′ phase [14]. The generalized stacking fault (GSF) surface determined from DFT calculation has been presented recently by D. Lv et al. [10]. The GSF calculations were conducted assuming stoichiometric $Ni_3Nb$ in the $D0_{22}$ crystal structure while focusing the faults on the (111) plane. The GSF results indicate that there exist multiple stable as well as unstable faults. The primary stable faults are the intrinsic stacking fault (ISF), superlattice intrinsic stacking fault (SISF), and a true antiphase boundary (APB). Unstable faults are also prevalent on the general stacking fault surface, and they come in the form of antiphase boundary-like (APB-like) and complex stacking fault (CSF) configurations. Stable configurations are defined as occurring at minima on the energy surface while unstable configurations fall on the slope of the energy surface. The atomic arrangement of these faults has been previously explored in detail by Lv et al. [10]. Major differences between the γ" ($D0_{22}$) and γ' ($L1_2$) fault structures can be attributed to the lower symmetry of the γ" phase and the corresponding GSF energy surface. For example, some $\frac{a}{2}\langle 110 \rangle$-type displacements can lead to an APB-like fault with lower symmetry compared to actual APBs. The corresponding displacement vectors will not end up at local minima of the GSF energy surface rendering the fault unstable so that it will naturally tend to relax into a lower-energy, stable configuration. Likewise, the CSF fault in the γ" lacks the three-fold axis symmetry present in an $L1_2$ CSF, which leads to an inherent mechanical instability. The intrinsic stacking fault (ISF) configuration is unique because it is stable and has a particularly low energy ( < 3 mJ/m$^2$ [10]). It is expected that the ISF configuration would form easily with the shear of a single $\frac{a}{6}[11\bar{2}]$ partial dislocation, but it cannot form directly from other Shockley partial displacements in the same plane. For further details, the reader is referred to the aforementioned publication by Lv et al. [10]. The ordering of the γ" phase when viewed on a $\langle 110 \rangle$ zone, and the displacement of the structure across an "edge-on" {111} shear plane for the several faults predicted by the GSF, are represented schematically in Figure 5.

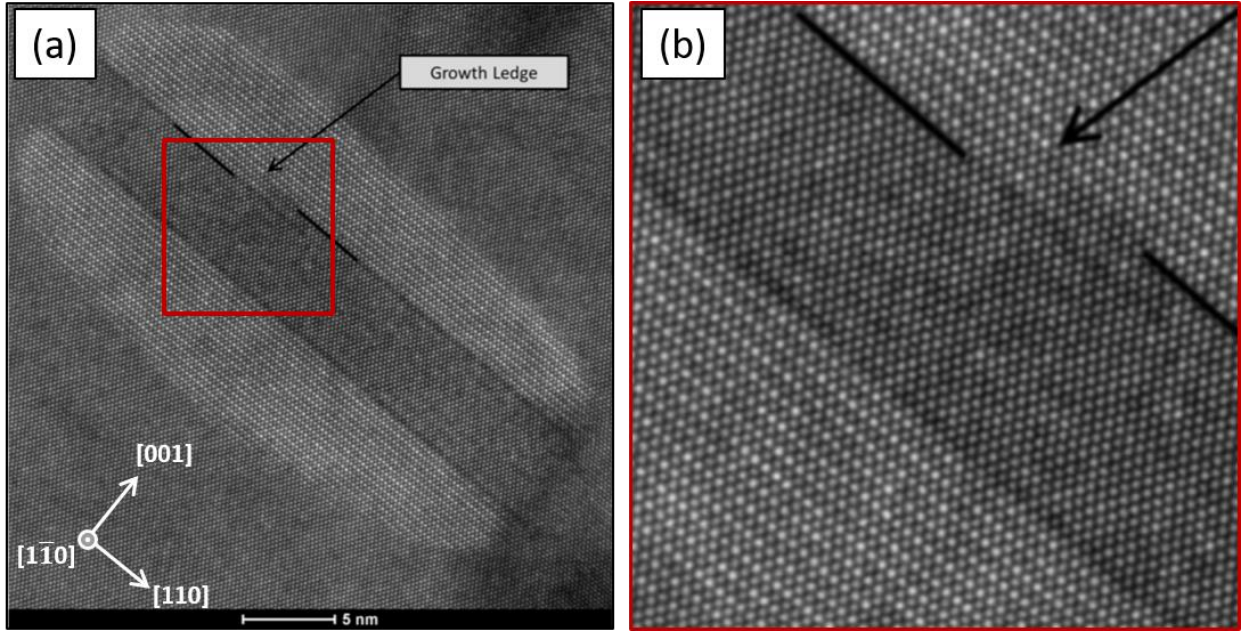

*Figure 4: (a) A growth ledge at the upper γ'/γ" interface as viewed along $[1\bar{1}0]$. The prominent (001) interface plane is edge-on. The magnification in (b) shows more clearly the ledge has a height equal to the γ" unit cell and doesn't extend to the opposite γ'/γ" interface. Growth ledges are rarely observed.*

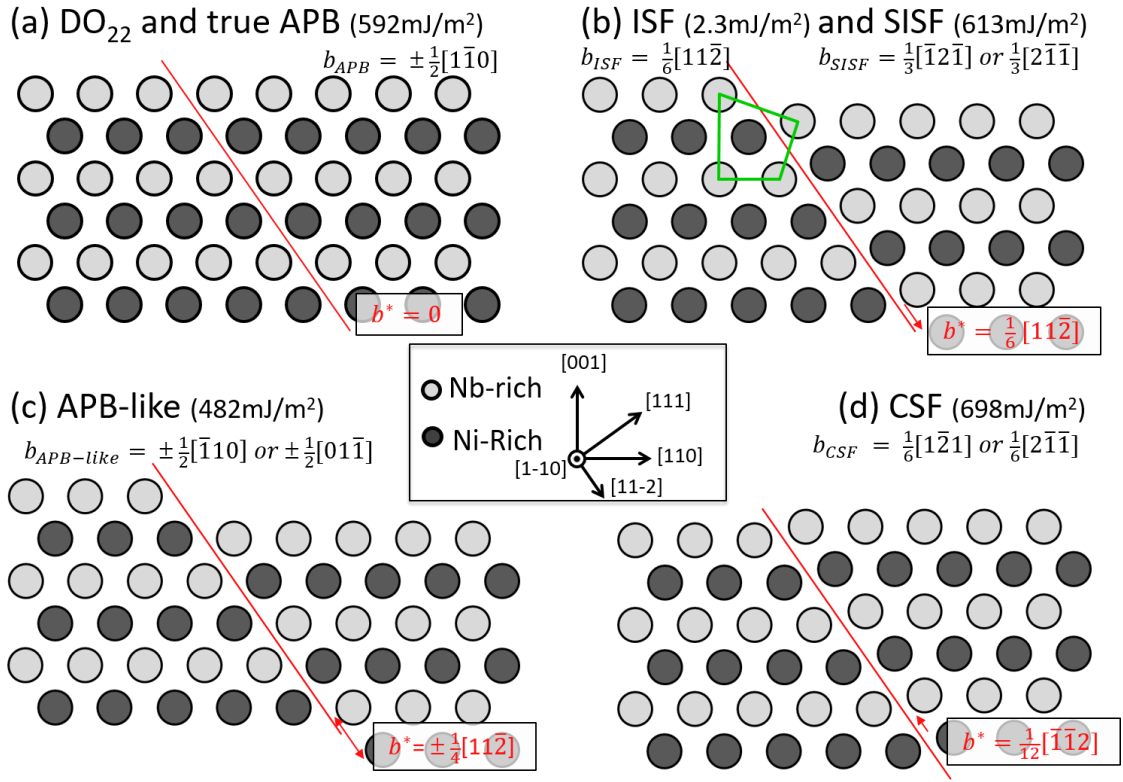

Figure 5: Schematic representations of the ordering contrast expected in atomic resolution HAADF STEM micrographs across the $(111)$ fault plane in a $[\bar{1}10]$ viewing direction. Planar fault energies from the GSF energy surface, total displacement vectors b within the $(111)$ plane and projected displacement vectors b* in the $(1\bar{1}0)$ plane of the paper are given for each major fault type a) APB, b) ISF and SISF, c) APB-like and d) CSF.

The $[\bar{1}10]$ viewing direction was chosen (wherein the c-axis of the selected γ" is parallel to the [001] crystal orientation) because the ordering of the γ" phase can be seen most readily when the (001) plane is viewed edge-on. This is the primary viewing direction for the HAADF imaging to be described below. A true APB would be indistinguishable from the perfect crystal structure since any shear would be parallel to the viewing direction, so the "in plane" structure and atomic ordering would not change, as shown in Figure 5a. SISF and ISF would be indistinguishable from each other, because their projected displacement vector b* in the plane of the paper is identical, as shown in Figure 5b. The ISF or SISF configuration can be identified in HAADF STEM images to be presented below by the polygonal structural unit created at the interface by the "bright" (high Z) Nb-rich atomic columns indicated in green in Figure 3. The CSF and APB-like sequences are uniquely identifiable with a single ⟨110⟩ view. A limitation of this method is that the γ' does not display strong ordering contrast for alloy 718, so APBs in the γ' cannot be readily distinguished from perfect crystal symmetry. Additionally, γ" does not show strong ordering contrast unless its (001) plane can be viewed edge-on, so the ordering in association with the fault can only be seen on the $[1\bar{1}0]$ zone for deformation on the (111) plane.

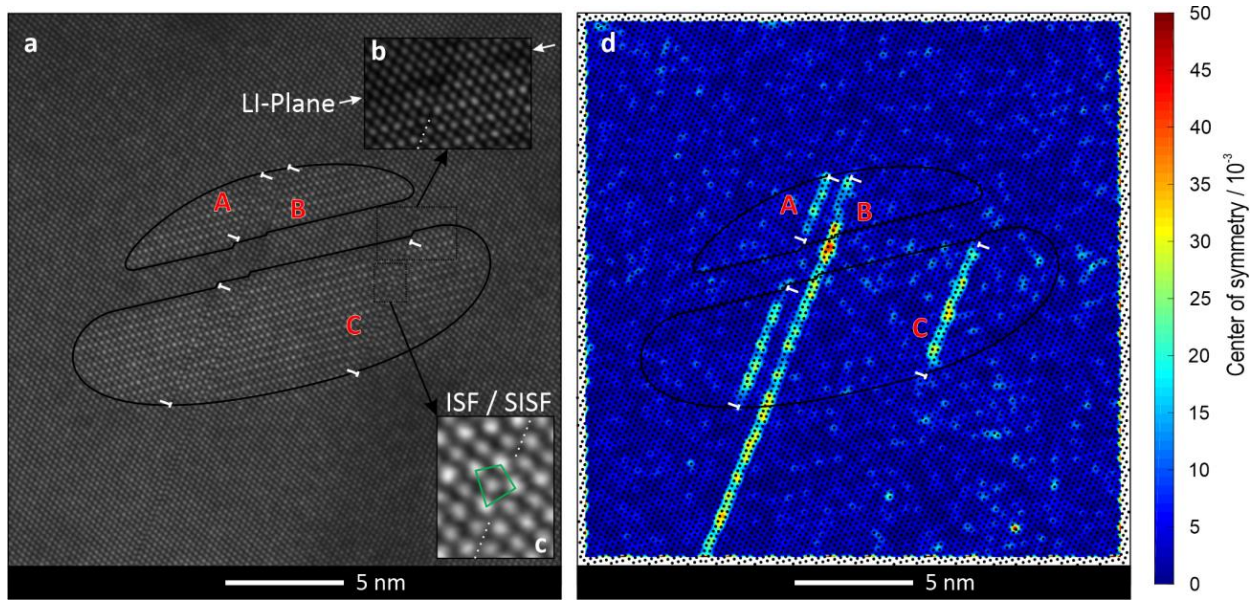

*Figure 6: (a) HAADF STEM of a γ'/γ" composite precipitate that has been sheared at 427°C on three different, parallel planes where shearing on two planes leave behind stacking faults in the γ" phase alone (events A and C), and one stacking fault extends through the γ', γ", and the matrix phases (shear event B). Arrows in (b) indicate the offset of the LI plane close to a stacking fault and (c) shows the polygonal structural unit at the stacking fault indicative of the ISF or SISF. The center of symmetry analysis [15] shown in subfigure (d) visualizes the faults more clearly.*

## 3.3 Zone Axis Analysis of Sheared Particles

Atomic resolution HAADF STEM microscopy was employed to validate the predictions of the GSF modeling results. Numerous stacking faults were found in association with the composite precipitates of samples deformed at 22 and 427°C. Stacking faults were generally constrained to the γ" phase. Partial dislocations were found at the γ/γ" interface. Only occasionally were stacking faults observed to extend into the matrix phase. When viewed on a ⟨110⟩ zone parallel to the fault plane, the local ordering of the fault was clearly evident and provided insight into the potential stacking fault configurations. Numerous stacking faults in multiple samples deformed at room temperature were investigated, and several examples of shearing events for composite particles with double-lobed γ" particle, containing a thin central γ' layer are discussed below.

As noted above, the LI plane at the γ'/γ" interface shown in Figure 3 can serve as an excellent marker that allows the shear history to be determined on a single {111} shear plane viewed edge-on. In Figure 6, there are three distinct shearing events occurring on parallel (111) planes that have been captured in the same particle, and we will now present an analysis of the far-left (shear event A) and middle (shear event B) structures since the faults and the off-set of the LI plane can be readily distinguished for these two events. The center of symmetry analysis shown in Figure 6d was conducted as described in [15] after refining atom column coordinates using Atomap [16].

### Shear event A: ISF$_{γ″}$/APB$_{γ′}$/ISF$_{γ″}$ Shearing

For shear event A shown in Figure 6 (left), stacking faults are clearly present in both γ″ lobes, while the central γ′ phase does not exhibit a fault and appears to be "perfect" registry. However, there is an offset of the LI plane at the γ″/γ′ interfaces which is equal to a full {111} d-spacing. This offset is consistent with a shear displacement on the (111) glide plane equal to full $a/2[1\bar{1}0]$ matrix dislocation; therefore, an APB must have been created in the γ′ phase. The stacking faults present in the γ″ lobes have a structure consistent with either an ISF or an SISF when viewed in the $[10\bar{1}]$ direction (compare Figure 5). Based on the fact that DFT calculations had shown that the energy of the ISF is dramatically lower than that of the SISF (see section 3.2 and reference [10]) and in agreement with the modelling presented later on in the paper (see section 4), it is reasonable to assume that the defect we observe is an ISF. Near both γ″/γ′ interfaces, partial dislocations must be present in order to accommodate the transition from the ISF in γ″ to the APB in γ′. Indeed, a Burgers circuit drawn around the lower γ″/γ′ interface (supplementary material) yields a closure failure equal to an $\frac{a}{6}\langle 112 \rangle$ Shockley partial in 30° orientation. The actual location of this partial dislocation core appears to be contained close to the interface within the γ″ phase. The ISF terminates at the lower γ″/γ interface with the presence of a bounding $\frac{a}{6}\langle 112 \rangle$ Shockley partial in 30° orientation, and with opposing sign to that just described. Thus, the ISF in the γ″ phase is bounded by a Shockley partial loop to accommodate the presence of the fault. Furthermore, these bounding Shockley partials produce a shear displacement that opposes the offset of the LI plane. Finally, the fault in the upper lobe exhibits the same attributes just described–an ISF that is isolated to the γ″ phase and bounded by a $\frac{a}{6}\langle 112 \rangle$ Shockley partial loop. An explanation for these features is offered below in Section 4, supported by phase field dislocation dynamics modeling.

Shear event C in Figure 6 only left behind a stacking fault exclusive to γ″, although, the presence of an APB in the adjacent γ′ cannot be ruled out. As the fault neither extends to γ nor γ′, as it is the case for shear event B described in the following section, it is assumed that a dislocation of the same Burgers vector as in shear event A is responsible for the creation of this fault. This would be in line with offset of the LI plane and the simulation results presented below. If this is true, an APB must indeed be present in the adjacent narrow γ′ "hamburger patty".

### Shear event B: ISF$_{γ″}$/SISF$_{γ′}$/ISF$_{γ″}$ Shearing

Shear event B in Fig. 6 (middle) differs from event A in several important respects. First, there is a continuous stacking fault that extends across all three phase regions. There is no discernable partial dislocation content at either γ″/γ′ interface. At the upper γ″/γ interface, a Shockley partial dislocation is present which terminates the stacking fault in the γ″ phase. In the case of the lower γ″/γ interface, the stacking fault extends into the matrix, and terminates some distance from the particle (not shown in the field of view in Fig. 5). Under the assumption that the faults in the γ″ phase are ISFs, the fault in the γ′ phase is expected to be a relatively low energy SISF. The plane of the supposed ISF in the upper γ″ lobe appears to shift to the adjacent {111} plane at the end of the fault. The reason for this shift is not presently understood, although it is noted that the lead configuration of $\frac{a}{3}\langle 112\rangle$ dislocations creating SISFs in the γ′ phase in several superalloys has been reported to occupy adjacent {111} planes, e.g. [15,17].

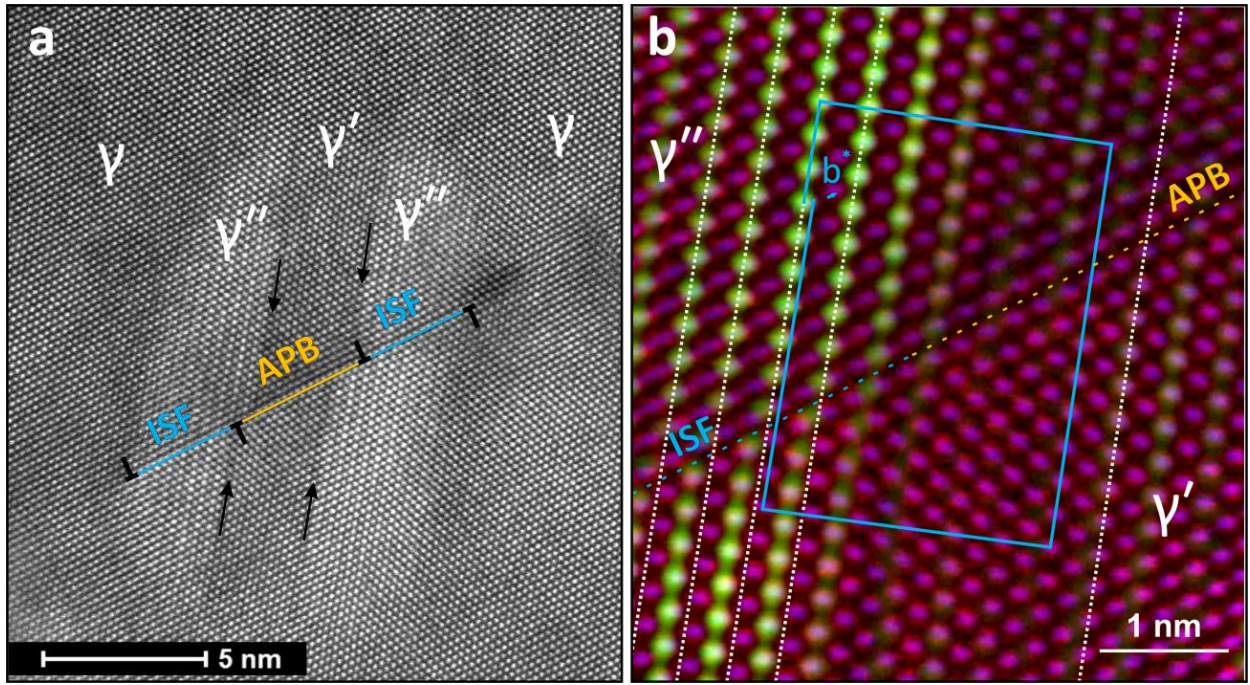

*Figure 7: A particle sheared at room temperature viewed down the $[\bar{1}10]$ zone axis and showing an ISF/APB/ISF fault configuration, but with an offset of the LI-plane (black arrows) that corresponds to shear by multiple $\frac{a}{2}\langle 110\rangle$ dislocations. As before, each γ″ lobe is enclosed in a Shockley partial dislocation loop; a) HAADF micrograph, b) Fourier-filtered composite image of the fault plane intersecting the left γ″/γ′ interface consisting of original HAADF (red), Nb-superlattice (green) and reconstructed lattice sites (blue).*

**Shearing by Multiple a/2<110> Dislocations:**

Numerous particles showed evidence of multiple shearing events. The offset of the LI plane is often greater than could be produced by a single matrix dislocation. Thus, multiple shearing events can occur on a single dislocation glide plane. It follows that the stacking faults and partial loops left behind after dislocation shearing does not completely inhibit the passage of further dislocations.

An example of such a case is shown in Figure 7a, which again shows a composite γ″/γ′/γ″ particle which has been sheared on a (111) plane after room temperature deformation. Note that the matrix does not exhibit any faults. However, it is clear in this case, that the co-precipitate has been sheared by multiple perfect $\frac{a}{2}\langle 110\rangle$ dislocations based upon the offset of the LI plane. In fact, the sheared configuration appears to be consistent with the passage of three $\frac{a}{2}\langle 110\rangle$ dislocations, with an additional partial displacement. Both of the γ″ lobes exhibit stacking faults (ISFs) that are bounded by opposing sign dislocations terminating the stacking faults near the γ″/γ′ and γ″/γ interfaces. The faults in both γ″ lobes and the central γ′ phase are not clearly distinguishable in this case, possibly due to variation of the shearing configuration through the thickness of the sample.

To reveal more detail, Figure 7b shows a higher magnification composite image of the left γ″/γ′ interface resulting from fast Fourier transformation (FFT) filtering. Common lattice sites and high-intensity Nb-rich sites have been visualized by introducing virtual apertures around the first order "fundamental spots" and "superlattice spots" in the FFT pattern, respectively, and reconstructing the image from them via inverse FFT. The reconstructed lattice sites (blue) and Nb-superlattice sites (green) have been superimposed on the original HAADF micrograph (red) to create the composite image. Even though the HAADF contrast around the fault is blurry, tracing the lattice planes across the fault (dashed white lines) reveals that γ″ contains an intrinsic stacking fault, while the γ′ phase appears to be in an APB configuration or perfect structure. Theoretical considerations about the faults – a group of dislocations that would result in a perfect structure in γ′ are not expected to form a stacking fault in γ″ – and the phase field simulations outlined below as well as those published in [18] indicate that the APB scenario is much more likely. Thus, the sheared structure exhibits the ISF $_{γ″}$/APB $_{γ′}$/ISF $_{γ″}$ structure; but it has been created by the passage of at least two $\frac{a}{2}\langle 110 \rangle$ dislocations.

# 4 Phase Field Simulation of Composite-Precipitate Shearing

It is difficult to rationalize the shearing process purely from the experimental observations, partly because of the complexity of the fault structures in the γ″ precipitates and the coupling between γ′ and γ″ precipitates and partly because of the ex-situ nature of the characterization. Moreover, the difficulty to distinguish SISF from ISF in the precipitates results in uncertainty in determining the deformation pathways. To gain a better understanding of the detailed deformation pathways involved in the dislocation shearing of γ′/γ″ composite particles, we use the microscopic phase field model (MPFM) to study the complete shearing processes of these particles, including all intermediate configurations [19,20]. MPFM has been applied to study shearing of γ′ and γ″ particles in Ni-base [10,18,21,22] and Co-base [23] superalloys. Part of the simulations outlined below have already been published (although with less detail) [18]. We include them here to be comprehensive.

Three types of typical dislocations and dislocation groups, with total Burgers vectors of $\frac{a}{2}\langle 110 \rangle$, $\frac{a}{2}\langle 112 \rangle$, and $\frac{3a}{2}\langle 110 \rangle$, are considered in this study and the results are compared with the experimental observations. The phase field model automatically includes dislocation-dislocation interaction and self-interaction. A hamburger-like γ″/γ′/γ″ composite particle, similar to those presented in Fig. 7 was used in the simulations, in which one of the three orientation variants of the γ″ precipitate is used without losing any generality. The glide plane in the simulations consists of 256×256 grid points, with a grid length of 0.25 nm. Periodic boundary conditions are used in all simulations. The applied shear stress is estimated from the yield strength of this alloy [24] and chosen to be 600 MPa. It is always applied along the total Burgers vector direction in the MPFM simulations unless specified otherwise. The dimension of the major-axis of the γ″ disk is estimated from Figure 4 and chosen to be 28nm. The lattice misfits between any two of the three phases are not considered in the following simulations since it would not alter the deformation mechanisms beyond exerting an additional resistance to the shearing process (i.e., altering the critical stress for shearing).

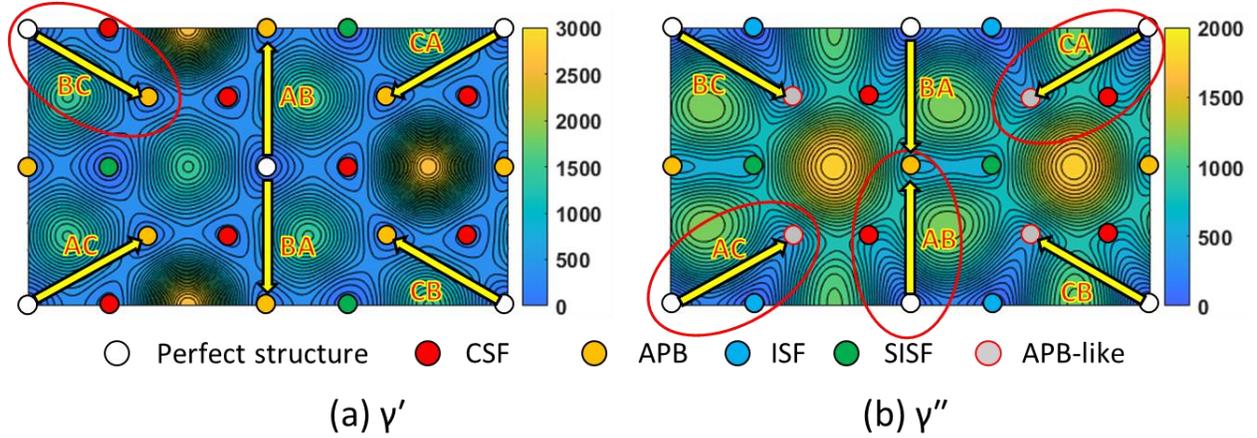

Figure 8: GSF energy (unit: mJ/m$^2$) surfaces of (a) γ' and (b) γ" phases obtained by DFT calculations [10,22]. The red circles indicate the independent deformation pathways of $\frac{a}{2}\langle 110 \rangle$ dislocations in these two crystal structures: all $\frac{a}{2}\langle 110 \rangle$ type dislocations have equivalent deformation pathways on the GSF energy surface of the γ' phase, whereas there are three distinct types on the GSF energy surface of the γ" phase. In the latter, AB is equivalent to BA, AC to BC, and CA to CB. Note that even though CSF and SISF exist in both phases, the fault energies are quite different.

## 4.1 Shearing of a Co-Precipitate by a Single a/2<110> Dislocation

There exist six possible $\frac{a}{2}\langle 110 \rangle$ Burgers vectors on any {111} plane in both γ' and γ" phases. Considering the symmetry of the γ' and γ" phases, all six Burgers vectors are equivalent in the γ' phase, whereas three of them are distinctly different in the γ" phase due to the loss of the threefold symmetry of the GSF energy surface on the γ" {111} plane, as shown in Figure 8. Therefore, AC, AB and CA dislocations (in Thompson notation) will behave differently when they interact with the co-precipitate and should be considered individually. In the simulations, the dislocations are assumed to glide from the bottom of the computational cell upward under the applied stress and interact with the precipitate that is located at the center of the cell.

Figure 9a(top row) shows the interaction between an AC ($\frac{a}{2}[01\bar{1}]$) dislocation and the co-precipitate that is outlined with the dashed yellow line in the background of frame one. The dislocation configurations (including core structures) are represented by the corresponding color contours in the GSF energy surfaces. The deep blue represents a perfect FCC stacking sequence unless specified otherwise, and the change of color from blue to green, then yellow and red, represents a higher GSF energy. In the γ' phase, the leading partial δC creates a CSF (Frame 1) and then the trailing partial Aδ transforms the CSF into an APB after shearing by the AC dislocation. In the γ" phase, due to the unstable APB-like fault that the AC dislocation would create, only the leading partial δC cuts into the γ" phase, forming a low-energy ISF, while the trailing partial Aδ loops around the γ" particle. The final sheared configuration is ISF$_{γ"}$/APB$_{γ'}$/ISF$_{γ"}$, with a Shockley partial Aδ looping around the γ" particles (Frame 4). This deformation microstructure corresponds to the shear event A observed in the experiment (Figure 6(a)). The deformation pathways of an AC dislocation in the γ′ and γ″ phases are shown on their GSF energy surfaces by green arrows in Figure 11(a) and (b), respectively. The expected offset of the LI plane is also visualized by the red arrow in Figure 11(b).

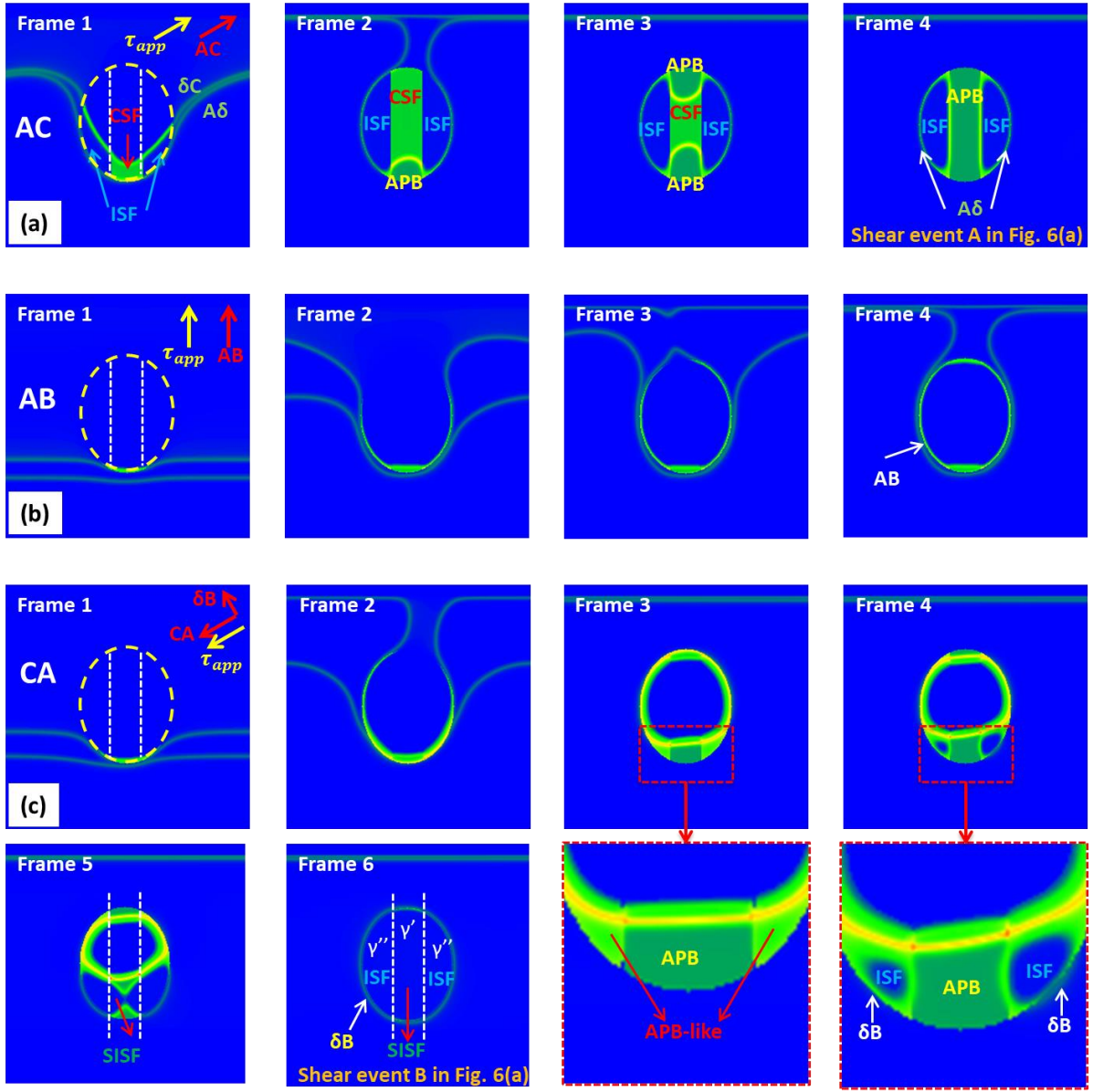

*Figure 9: Interaction of individual (a) AC, (b) AB, and (c) CA dislocation with a hamburger-like co-precipitate on the (111) glide plane. The precipitates are outlined by the dashed white and yellow lines in Frame 1. The applied stress direction is along the total burgers vector of the shearing dislocation.*

Figure 9b (second row) shows shearing of the co-precipitate by an AB dislocation. Due to the high APB energy in the γ″ phase, the entire co-precipitate is looped by the AB dislocation, even though the AB dislocation could create an APB in the γ′ phase if it was an isolated γ′ precipitate or a single-lobed instead of a dual-lobed co-precipitate.

Figure 9c (third and fourth rows) shows shearing of the co-precipitate by a CA dislocation. The two partials shear together as a compact dislocation. They create an APB in the γ′ phase and an APB-like fault in the γ″ phase (inset of Frame 3). A Shockley partial δB is then nucleated in the γ″ phase, transforming the APB-like fault into an ISF (inset of Frame 4) because the APB-like fault is unstable [10]. Then δB propagates into the γ′ phase, transforming the APB into an SISF (Frame 4 and 5). The final sheared configuration is ISF$_{γ″}$/SISF$_{γ′}$/ISF$_{γ″}$, with a Shockley partial δB looping around the whole co-precipitate, which agrees well with the experimental observations (shearing event B in Figure 6(a)). However, unlike the experiment where the extended fault ISF$_{γ″}$/SISF$_{γ′}$/ISF$_{γ″}$ expands further into the γ matrix, the remnant Shockley partial looping around the whole co-precipitate does not expand into the γ phase in the simulation. This is because the applied stress is perpendicular to the Burgers vector of the remnant Shockley partial (Frame 1 of Fig. 9c). If the applied stress is, for example, along the CA+CB ($[\bar{1}\bar{1}2]$) direction, the remnant Shockley partial will further expand into the γ matrix, as indicated by the simulation result shown in Fig. 10, where the yellow dash line indicates the matrix-precipitate interface. The deformation pathways in the γ′ and γ″ phases are shown on their GSF energy surfaces by the yellow arrows in Figure 11a and b, respectively. The expected offset of the LI plane is also labeled in the red arrow in Figure 11b. This is a clear result of the interplay between γ′ and γ″ phases in the dual-lobed co-precipitate because such a deformation pathway would not occur if the precipitate is monolithic or single-lobed.

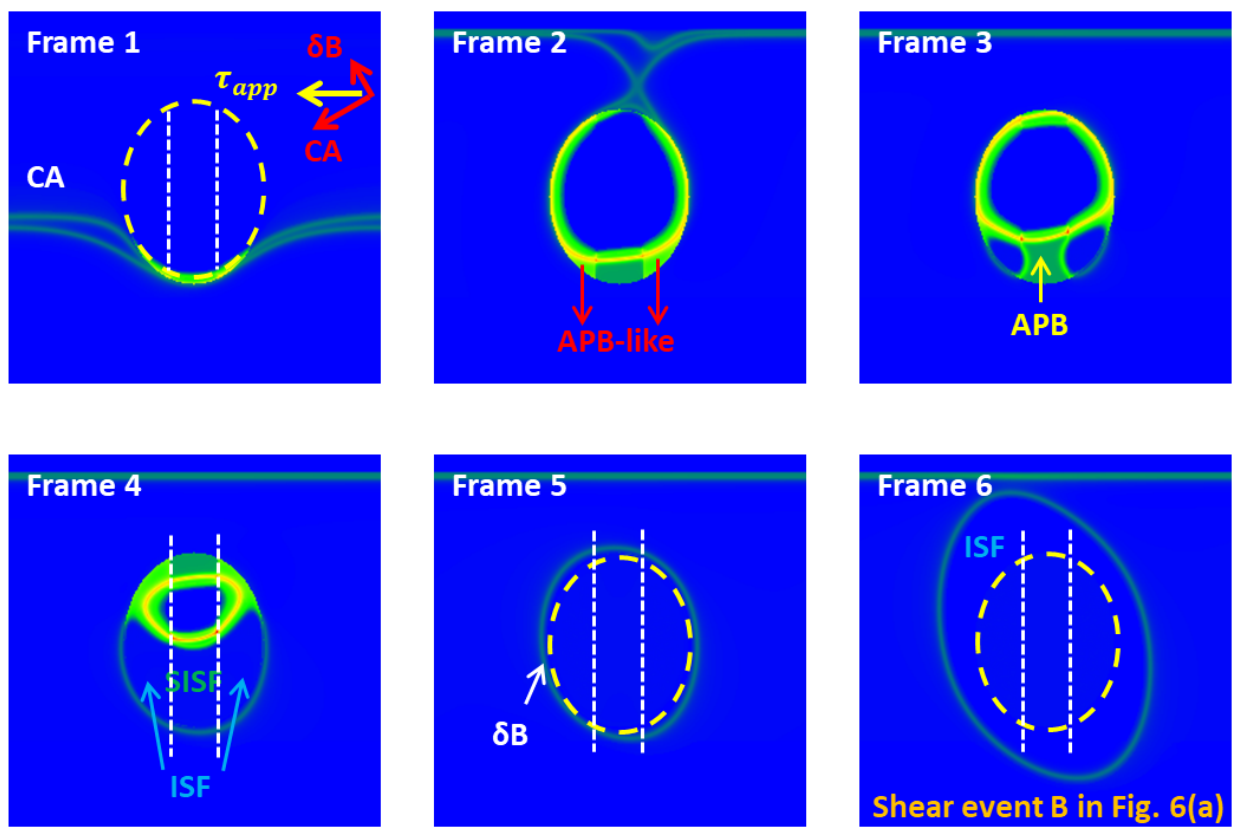

*Figure 10: Interaction between CA dislocation and the hamburger-like co-precipitate under an applied stress along CA+CB ($[\bar{1}\bar{1}2]$) direction. The dashed yellow line indicates the shape of the co-precipitate. Contrary to Fig. 9c, the nucleated Shockley partial loop is expanding, because the stress direction drives its expansion, consistent with the experimentally observed shear event B shown in Fig. 6(a).*

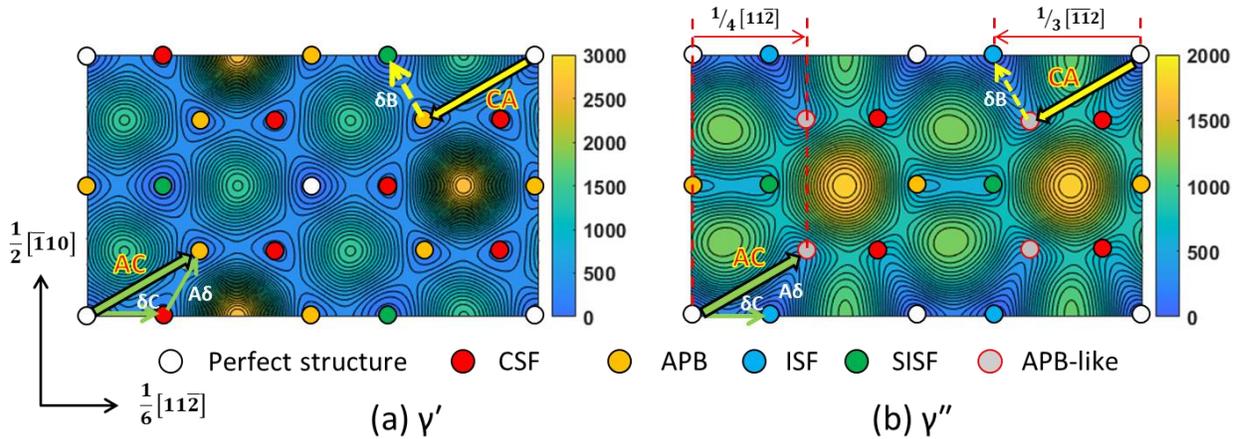

Figure 11: Deformation pathways of AC and CA dislocations in (a) γ' and (b) γ'' phases shown in Fig. 9. The deformation pathway of the AB dislocation is not shown because, instead of shearing, it loops around the co-precipitate.

### 4.2 Shearing of a Co-Precipitate by a/2<112> Dislocation Groups

Figure 12 (top row) shows the interaction between an AC+AB ($\frac{a}{2}[\bar{1}2\bar{1}]$) dislocation group and the co-precipitate. In the γ' phase, the AC dislocation first creates an APB (inset of Frame 1) after which an SISF is formed by the leading partial δB of the AB dislocation (inset of Frame 2). The trailing partial Aδ of AB finally cuts into the γ' phase, leaving an APB behind after shearing of the $\frac{a}{2}\langle 112 \rangle$ dislocation group into the particle (inset of Frame 3). In the γ'' phase, an ISF is created by the leading partial δC of AC (inset of Frame 1) and an SISF ribbon forms after the leading partial δB of the AB dislocation enters the particle (inset of Frame 3). The trailing partial Aδ of AB finally transforms the SISF into the unstable APB-like fault, which subsequently reacts to an ISF under the formation of a δB Shockley partial loop which is left behind around the γ" lobes (Frame 4 and 5). Unlike the remnant Shockley partial created by the CA ($\frac{a}{2}[0\bar{1}1]$) dislocation, this Shockley partial δB does not propagate into the γ' phase due to the high energy of the CSF it would introduce, even though the applied stress does drive the expansion of δB. However, it does slightly extend into the matrix phase, as indicated by the dashed yellow line in Frame 5. The deformation pathways in the γ' and γ" phases are indicated by the orange arrows in Figure 13a and b, respectively. As will be discussed further below, this predicted shearing configuration is similar to the experimental observation shown in Figure 7, although, the corresponding offset of the LI plane (red arrow in Figure 13b) would be less than what is observed experimentally.

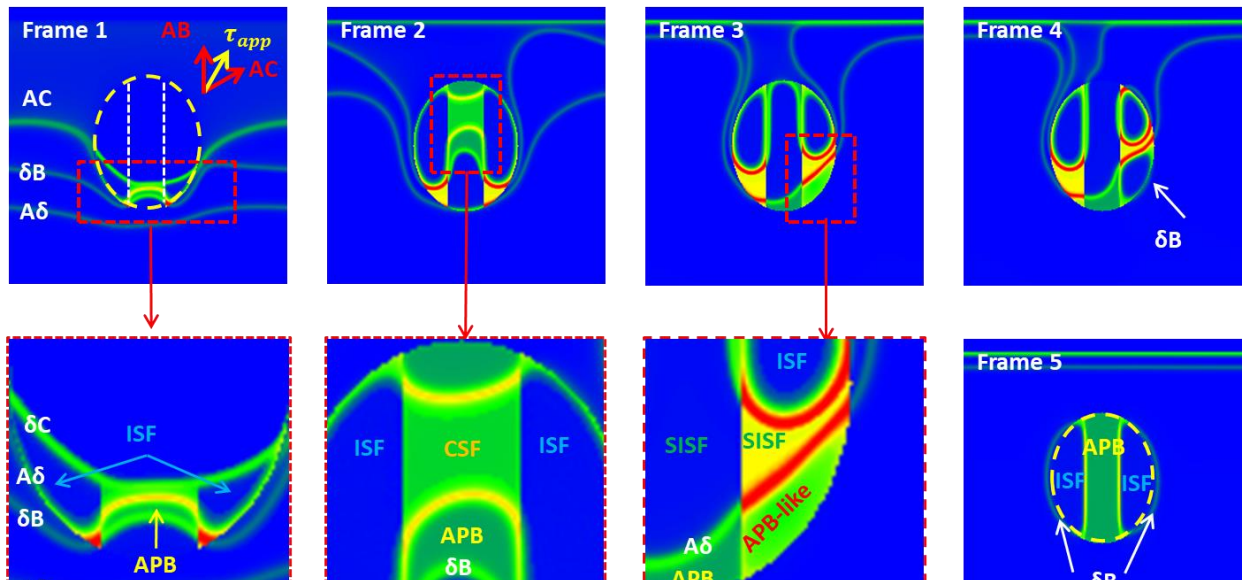

Figure 12: Interaction between AC+AB dislocation (i.e. $\frac{a}{2}[\bar{1}2\bar{1}]$) and the dual-lobed co-precipitate.

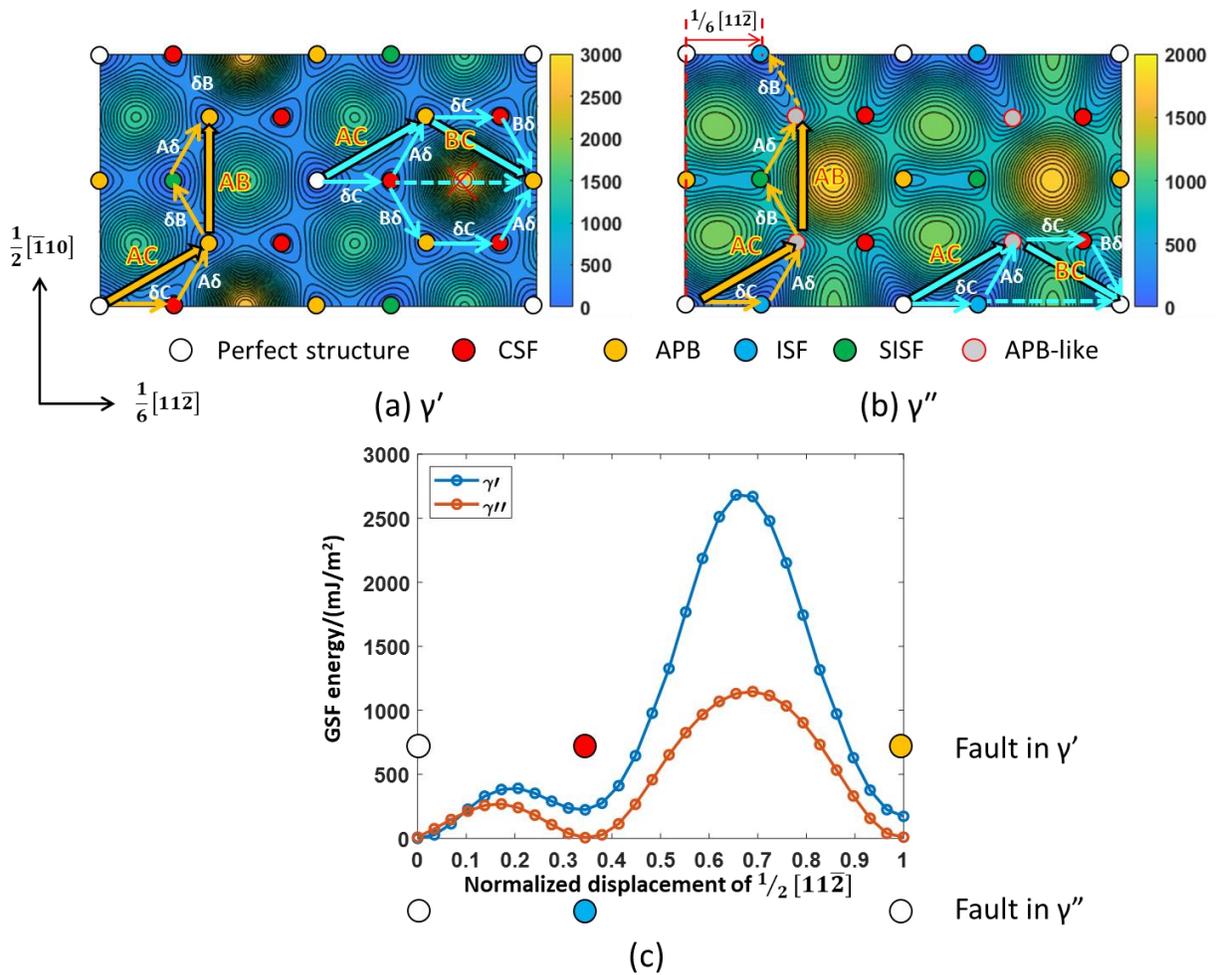

Figure 13: Deformation pathways of AC+AB and AC+BC dislocation groups in (a) γ' and (b) γ" phases. (c) Minimal energy pathway along the displacement of $\frac{a}{2}[11\bar{2}]$ (dashed blue arrow) on the GSF of γ' and γ" phases.

Figure 14 shows the interaction between the co-precipitate and another $\frac{a}{2}\langle 112 \rangle$ dislocation group, AC+BC. The dashed yellow line indicates the matrix-precipitate interface. The leading partial δC of AC cuts the co-precipitate and creates an ISF in the γ″ phase and a CSF in the γ′ phase (Frame 2). Then, the remaining three Shockley partials cut into the γ″ particle as a compact dislocation and restore the perfect structure in the γ″ phase (Frame 3). After the compact dislocation shears through γ″ and loops around γ′, the trailing partials Aδ (of AC) and Bδ (of BC) cut into the γ′ phase from the top (the rear side) left and right sides of the γ′ precipitate (inset of Frame 5), followed by the remaining two leading partials (insets of Frame 6). The final configuration is APB in the γ′ phase and perfect structure in the γ″ phase, with no residual dislocation content (Frame 7). The deformation pathways in the γ′ and γ″ phases are shown by the blue arrows in Figure 13a and b, respectively. The asymmetric cutting (i.e., different Shockley partials cutting into the γ′ from different sides, inset of Frame 5) is caused by the asymmetric dissociation of the compact dislocation (Frame 3 in Figure 14), where Aδ leads on the left while Bδ leads on the right. The compact dislocation has a total Burgers vector of $\frac{a}{3}[11\bar{2}]$, as shown in Figure 13 with a dashed blue arrow. Partial Aδ of AC and Bδ of BC are symmetric with respect to the $[11\bar{2}]$ direction. With the applied stress also being parallel to $[11\bar{2}]$ direction, the asymmetric dissociation of the compact dislocation is caused solely by the Poisson's effect. The edge dislocation has a larger line energy as compared to its screw counterpart due to the non-zero Poisson's ratio and, thus, its length shortens faster. With the compact dislocation looping around the γ′ particle, partial Aδ of AC has more edge component on the left side while partial Bδ of BC has more edge component on the right side. If we set the Poisson's ratio to zero in the simulation, this asymmetric dissociation disappears, as indicated in Appendix A,

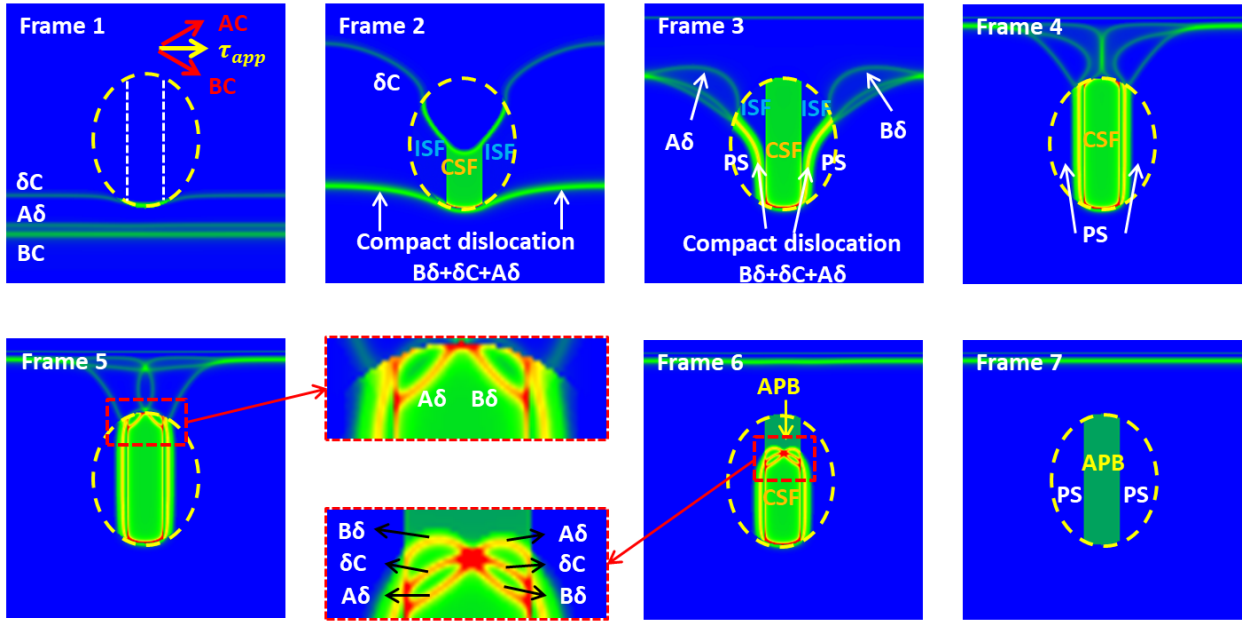

Figure 14: Interaction between AC+BC dislocation group (i.e., $\frac{a}{2}[11\bar{2}]$) and the hamburger-like co-precipitate.

assuming the leading partial δC of AC has passed).

Another interesting feature is that the compact dislocation cannot directly transform the CSF into an APB by cutting into the γ' phase from the side it approaches it from (bottom) because it would force the deformation pathway to pass the global energy maximum on the GSF surface of the γ' phase (see the dashed blue arrow in Figure 13a). Figure 13c compares the energy landscapes in the γ' and γ" phases along the displacement of $\frac{a}{2}[11\bar{2}]$ (dashed blue arrows in subfigures a and b). The energy along this pathway for the γ' phase is much higher than that for γ" phase. Thus, when the compact dislocation enters the γ" phase and restores the D0$_{22}$ structure, it gets stuck at the lower γ/γ' interface (Frame 3). On the top, due to the asymmetric dissociation of the compact dislocation in the matrix, the three partials Aδ, δC and Bδ are no longer compact any more after they loop around the γ' particle. Partial Aδ of AC and partial Bδ of BC can easily cut into the γ' particle and transform the high-energy CSF into APB, with a dislocation node in the γ' phase (inset of Frame 6 in Figure 14). This is another case where the interplay between γ' and γ" phases in the dual-lobed co-precipitate that results in unique dislocation configuration and addition shear resistance in the shearing process.

### 4.3 Shearing of the Co-Precipitate by 3 Consecutive CA Dislocations

Even though a group of AC+AB dislocations can create the experimentally observed ISF$_{γ"}$/SISF$_{γ'}$/ISF$_{γ"}$ configuration shown in Figure 7, it must be pointed out that since the AB displacement is parallel to the edge-on viewing direction, it is not possible to draw conclusions about the number of AB dislocations that have sheared the co-precipitate, because the LI plane will not be offset [4]. Yet, the distinct LI plane indicates a displacement of more than one $\frac{a}{2}\langle 110 \rangle$ dislocation. Another group of dislocations other than AC+AB must be responsible for the ISF$_{γ"}$/SISF$_{γ'}$/ISF$_{γ"}$ configuration shown in Figure 7. We propose a scenario where three consecutive CA dislocations shear the co-precipitate.

Figure 15 shows the phase field simulation results for 3 consecutive CA dislocations interacting with the co-precipitate. As described above (Figure 9c), the first CA leaves behind an extended fault in the whole precipitate with SISF in the γ' phase and ISF in the γ" phase (Frame 5), also creating a remnant Shockley partial δB in the γ" phase first and then extend into the γ' phase. In the γ' phase, the trailing partial Cδ of second CA enters the precipitate and forms APB (Frame 5 and 6). The leading partial δA then cut in the precipitate followed by the remnant Shockley partial δB being pushed back in (Bδ) and form the perfect structure in the γ' phase (Frame 6). The third CA then cuts the γ' phase and creates an APB as the final configuration. The deformation pathway on the GSF energy surface of γ' is shown in Figure 16a. In the γ" phase, the trailing partial Cδ of the second CA restores the perfect structure with the leading partial δA and remnant Shockley partial δB looping the γ" precipitate (Frame 6). When the third CA cuts in, an APB-like structure is formed by reacting with the residual δA and δB loops (Frame 7 and 8), which instantaneously decomposes into an ISF and another remnant Shockey partial δB loop around the γ" precipitate (Frame 9 and 10). However, the cutting sequence of the third CA and the previous δA and δB are different for the two individual γ" precipitates. For the γ" precipitate on the right, the previous δA enters the γ" first and then the previous δB is pushed in (Bδ on the GSF), which leads to the creation of an APB (inset of Frame 7). The third CA then enters γ" and forms unstable APB-like, which transforms into ISF by nucleating another remnant Shockley partial δB. The sequence in which the dislocations enter γ" is the same as that of γ'. For the γ" precipitate on the left, after the previous δA enters γ", the third CA cuts in and forms SISF, with the remnant partial δB being pushed inside γ", forming an unstable APB-like fault, which instantaneously transforms into an ISF by nucleating another remnant Shockley partial δB (inset of Frame 8). The sequence in which the dislocations enter γ" is different from that of γ', leading to a dislocation node forming at the γ'/γ" interface, shown in the inset of Frame 8. The deformation pathway on the GSF energy surface is shown in Figure 16b, wherein the expected offset of the LI plane is also labeled in the red arrow.

---

[4] Due to the orientation relationship between γ' and γ", and the reduced symmetry of the γ", any dislocation that would create a true APB in γ" from the perfect structure will have a Burgers vector parallel to the γ'/γ" interface and will thus never create a discernible offset of the LI plane.

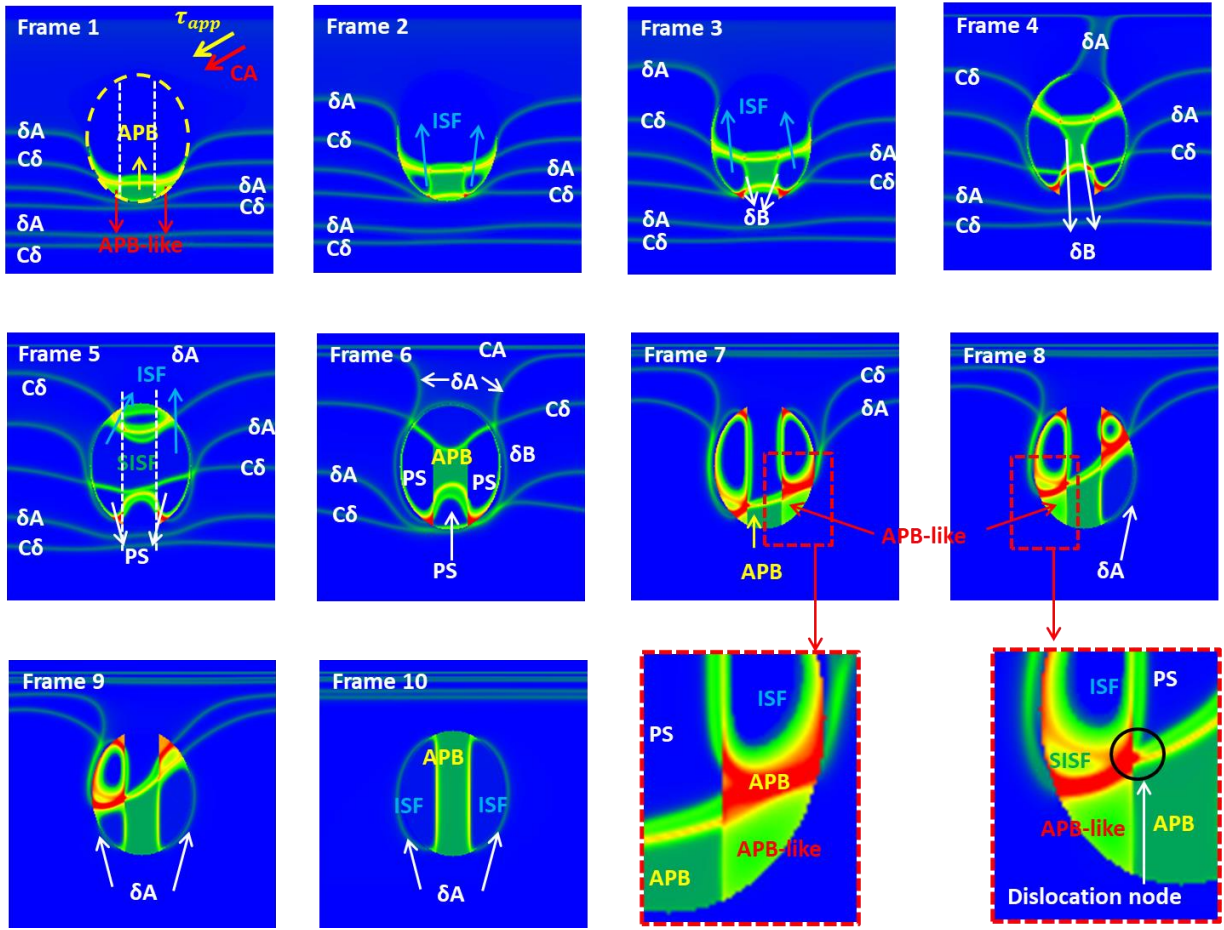

*Figure 15: Interaction between three CA dislocations and the γ"/γ'/γ" co-precipitate.*

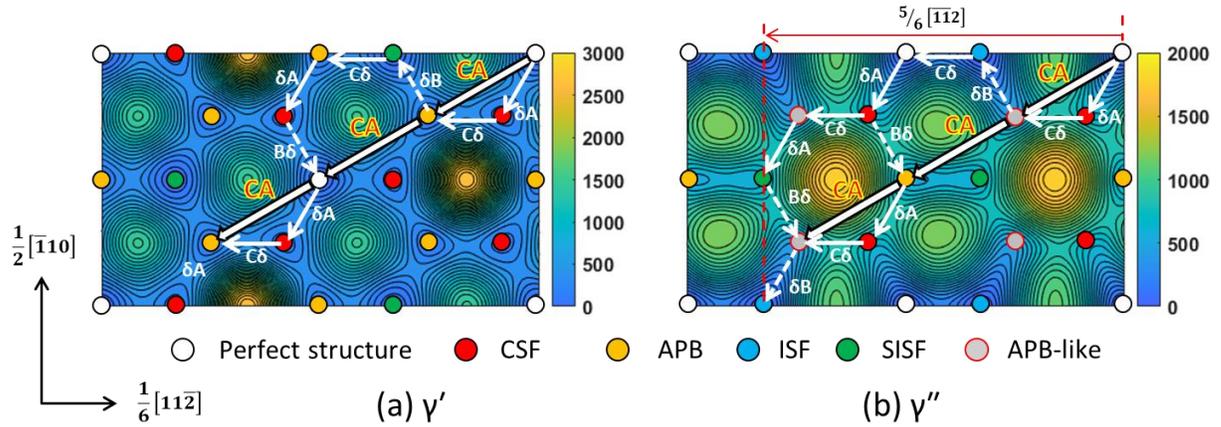

*Figure 16: Deformation pathways of three consecutive CA dislocations in (a) γ' and (b) γ'' phases.*

This is another important result where the interplay between γ' and γ'' influences the deformation pathway. Note that the deformation process of the γ' phase for the first two CA dislocations should be equivalent to the typical $a\langle 110\rangle$ dislocation shearing (i.e., "APB shearing") of a monolithic γ' precipitate. In fact, Figure 15 shows that the third CA dislocation is not strongly coupled with the previous two (Frame 1-6). However, unlike the usual APB ribbon associated with an $a\langle 110\rangle$ dislocation, additional APB + SISF faults are created (Figure 15 - Frame 5 and Figure 16a), which should yield a higher shear resistance and hence a larger strengthening effect. This is due to the existence of the γ'', which forms an ISF after the first CA dislocation passed, with details shown in Figure 9c. In turn, the γ' phase also affects the deformation pathway of the γ'' phase. As mentioned above, when the third CA shears the co-precipitate, it exhibits an asymmetric deformation pathway for the two sides of the γ''/γ'/γ'' co-precipitate. If we just focus on the faults that are created within the γ'' phase, the CSF → APB-like → SISF → APB-like pathway (γ'' precipitate on the left, inset of Frame 8) has a lower energy (2276 mJ/m$^2$) than that (2470 mJ/m$^2$) of the CSF → APB → CSF → APB-like pathway (γ'' precipitate on the right, inset of Frame 7). This suggests that the former pathway is favored by the γ'' precipitate. But the γ' phase prefers the latter and therefore forces this pathway. A reference simulation is done removing the γ' phase to see how it influences the deformation process of γ'', shown in Appendix B. Without the γ' phase, the γ'' phase prefers the formation of SISF (Frame 8), i.e. the deformation pathway for the third CA is CSF → APB-like → SISF → APB-like, which has a lower energy pathway compared to the other one. Moreover, since the applied stress is perpendicular to the last partial δB, a stable SISF fault is created within the γ'' precipitate despite a relatively high SISF energy compared to its counterpart in the γ' phase. While in the previous simulation with the γ' patty in the middle, the asymmetry of the cutting sequence prohibits the separation of these partials. Therefore, there will be no occasion where a single δB is cutting the precipitate and a group of dislocations moving together is always favored. There is clearly a synergistic interplay between the deformation pathways in the γ' and γ'' phases in a co-precipitate configuration, which forces high-energy pathways that would not be present for individual monolithic precipitate and therefore provides an additional strengthening effect.

## 5 Comparison of Experiments with Simulations of Co-Precipitate Shearing

In the following, we compare the findings from the experiments with what is found by the simulations. A summary of the results with references to the respective figures is shown in Figure 17. As before, the discussion below assumes shearing of one γ″/γ′/γ″ co-precipitate with a $[001]_{\gamma'} \parallel [001]_{\gamma''}$ & $(110)_{\gamma'} \parallel (110)_{\gamma''}$ orientation relationship on the (111) plane (ABC in the Thompson notation), but it is generally applicable to all three orientation variants and four {111} slip planes.

### 5.1 Fault configurations

With respect to the planar faults left behind in the γ′ and γ″ phases, two different configurations have been identified experimentally for γ″/γ′/γ″ co-precipitates: $ISF_{\gamma''}/APB_{\gamma'}/ISF_{\gamma''}$ and $ISF_{\gamma''}/SISF_{\gamma'}/ISF_{\gamma''}$, see Figure 17a and b. The formation of these faults is confirmed in the simulations. As shown in Figure 9a, the $ISF_{\gamma''}/APB_{\gamma'}/ISF_{\gamma''}$ configuration is created through shearing by a $\frac{1}{2}[01\bar{1}]$ dislocation (AC). The trailing $\frac{1}{6}[\bar{1}2\bar{1}]$ partial (Aδ) doesn't cut into the γ″ because it would transform the low-energy ISF into an unstable and high energy APB-like fault. Thus, Aδ loops around both γ″ lobes are left behind. The two sides of these loops appear as edge-on opposite sign partials enclosing an ISF in the HAADF micrographs (compare Figure 17d and Figure 6).

The ISF/APB/ISF fault configuration depicted in Figure 17b) and e) is created when a $\frac{1}{2}[0\bar{1}1]$ dislocation (CA) shears the co-precipitate as shown in Figure 9c. If the stress direction favors the expansion of this ring, it can also move on into the γ matrix (see Figure 9c vs. Figure 10). This corresponds well to the experimental results shown in Figure 6 (shear event B), where one side of the loop is still pinned at the γ/γ″ interface while the other one has expanded into γ, creating a stacking fault there as well.

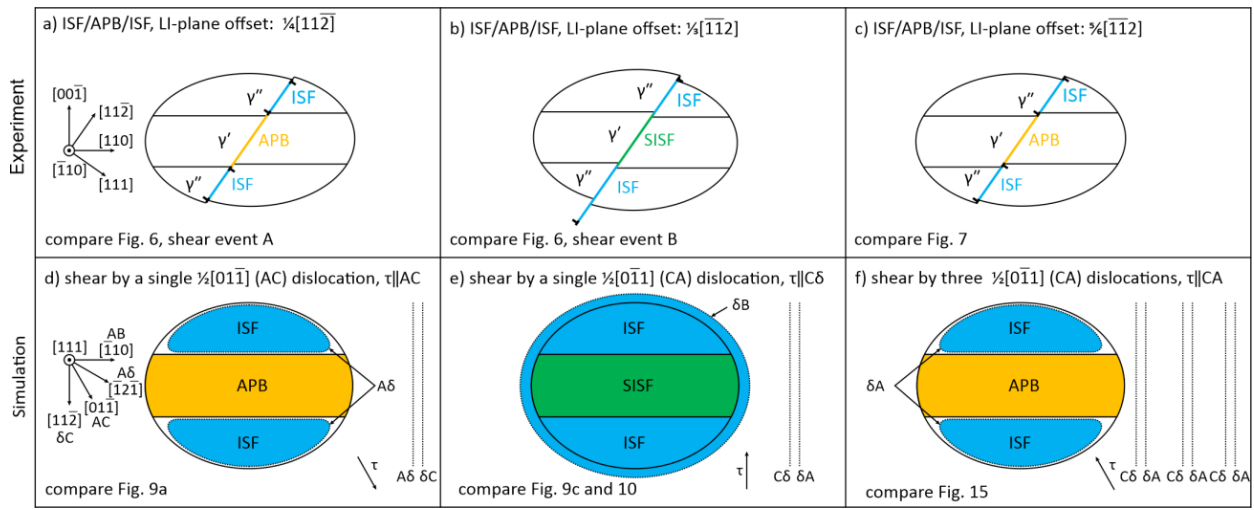

*Figure 17: Summary and comparison of experimental (a-c) and simulated (d-f) fault structures with references to the corresponding figures. Experimentally determined and simulated displacement vectors (compare red arrows in Fig. 11, 13 and 16) correspond very well.*

It is worth noticing that a deformation pathway that is identified in the simulations has not yet been observed experimentally. Figure 9b shows that when an $\frac{1}{2}[\bar{1}10]$ dislocation (AB) approaches the co-precipitate, both partials bypass it and leave a loop around it. Although there is no experimental evidence for this configuration, its occurrence is likely. As there would be no stacking fault involved and no offset of the LI-plane, this structure would be extremely difficult to identify in high-resolution HAADF-imaging.

The three shear directions considered in the simulations are representative of all shearing events through the same $\frac{a}{2}\langle 110 \rangle$ dislocations on any given {111} plane, as have been discussed earlier in Section 4.1 (see Fig. 8).

## 5.2 Shearing by multiple dislocations

The ISF$_{\gamma''}$/APB$_{\gamma'}$/ISF$_{\gamma''}$ fault configuration depicted in Figure 17c is the same as the one shown in subfigure a, but with a larger offset of the LI plane, indicating that more than one $\frac{a}{2}\langle 110 \rangle$ must have passed through the co-precipitate. However, pairs of like dislocations would restore both phases to their perfect structure some of which leaving dislocations loops behind. Hence, it is possible that unlike dislocations are responsible for the observed fault configuration. One such possibility was presented in Figure 12: Shearing by a pair of a $\frac{a}{2}[01\bar{1}]$ and a $\frac{a}{2}[\bar{1}10]$ (AC+AB) dislocation. Co-planar $\frac{a}{2}\langle 110 \rangle$ dislocations with different Burgers vector have been reported to be responsible for the formation of stacking fault ribbons in the low-temperature/high-stress regime of SX Ni-base superalloys, e.g. Wu et al. in [25], which also contains a historical overview over previously conceived a $\langle 112 \rangle$ stacking fault shearing mechanisms. In a prior work, we also found evidence of $\langle 112 \rangle$ dislocation activity in IN718 [11]. In the present case, similar to the case of shearing by a single AC dislocation, a partial dislocation loop is nucleated in the last step to transform an APB-like fault in the γ″ into an ISF. These two dislocations would not create a large enough offset of the LI plane either, though. In fact, the observable displacement of AC+AB shearing is the same as in the case of a single AC dislocation as the AB displacement is parallel to the TEM viewing direction. This is not only a problem of the chosen viewing direction: generally, the Burgers vectors of AB-type dislocations (those that would create a true APB in the γ″) are always parallel to the γ′/γ″ interface and would thus never create an observable offset of the LI-plane.

To create a larger offset of the LI-plane, and still end up with the ISF$_{\gamma''}$/APB$_{\gamma'}$/ISF$_{\gamma''}$ configuration, more dislocations that have a component in $[\bar{2}11]$ direction (e.g. ±AC or ±BC) are required. One such scenario was presented in Figure 15 (compare also Figure 17f): Shearing by three CA dislocations. According to the phase field simulations, this configuration matches not only the experimentally observed fault configuration (ISF$_{\gamma''}$/APB$_{\gamma'}$/ISF$_{\gamma''}$) but also the associated displacement of the γ′/γ″ interfaces (LI-plane), namely the equivalent of three full and one partial dislocations.

More possibilities exist that partially involve activity of unlike $\frac{a}{2}[01\bar{1}]$ dislocations on the same slip plane. Given that the formation of the experimentally observed configuration can be explained without the activity of unlike dislocations, it seems the more likely scenario.

# 6 Summary and Conclusions

Several different fault configurations were identified for γ″/γ′/ γ″ co-precipitates in IN718 at low temperatures. The atomically flat "low-intensity (LI) plane" in HAADF micrographs of γ″/γ′ interfaces, caused by an enrichment of Al, was introduced as a useful marker to determine the shear history of such a composite particle, even in the absence of dislocations or stacking faults. One weakness of this method is that one of the three shear directions in any given {111} plane cannot be traced, because its displacement vector is parallel to the γ″/γ′ interface and would not create an offset of the LI plane. Thus, possible participations of such dislocations (such as AB and BA on (111) in the Thompson notation) in any of the deformation pathways cannot be ruled out with the method presented herein.

It was shown that multiple deformation mechanisms can operate at the same time, even in the same co-precipitate. Density-functional-theory-informed phase field simulations were shown to be a useful tool to model the deformation mechanisms likely to be operative. All fault configurations experimentally observed have been predicted by the simulations. The simulations also found a mechanism (shear by a single AB or BA dislocation) that might be operative but would be hard to identify experimentally because it does not leave planar faults behind so that a spot on dislocation on both sides of the co-precipitate in the HAADF micrograph would be the only indication of it.

Due to the lower symmetry of the γ″ phase, there are planar fault configurations that do not exist in the γ′ phase, e.g., an ISF and an APB-like configuration. Most of the deformation mechanisms discussed herein are dominated by the fact that the ISF in γ″ has an ultralow energy (~3mJ/m$^2$) and that the APB-like fault is unstable. The most important thing to note, however, is that the deformation pathway of the hamburger-like co-precipitate is different from the ones of either precipitate in its monolithic form (or even from single-lobed co-preciptates), as the presence of the respective other phase constrains the number of energetically favorable mechanisms. This is generally associated with an increased shear resistance.

One of the planar fault configurations in the co-precipitate exhibits an offset of the LI plane that can only be explained by shear by dislocation groups of at least three individual $\frac{a}{2}\langle 110 \rangle$ dislocations. The activity of further unlike dislocations cannot be ruled out but is not required.

The shear through quadruplets of like $\frac{a}{2}\langle 110 \rangle$ dislocations, previously hypothesized necessary in order to restore γ″ to its perfect structure [4], was never observed in our experiments. In fact, the DFT-informed GSE surfaces and the corresponding phase-field simulations indicate that other mechanisms might be energetically more favorable.

Contrary to the room temperature experiments, stacking faults were regularly found to extend from the γ″ lobes into the γ matrix phase at elevated temperatures. Possible explanations for this include an inherently decreasing γ stacking fault energy, accelerated elemental segregation to the faults and a decrease of the coherency stress that might be pinning the partials at the γ/γ″ interfaces at room temperature. These stacking faults might facilitate microtwinning with ongoing deformation.

**Acknowledgments**

Financial support for DM was provided by the Materials Affordability Initiative GE-10, while MJM and YW acknowledge NSF DMREF under grant number 1922239 for continued support. CHZ acknowledges financial support from the Alexander-von-Humboldt Foundation. The authors would like to thank Dr. Feng Shen for his assistance with EDS mapping.

# Appendix

## A. Interaction between AC+BC dislocation and the hamburger-like co-precipitate with Poisson's ratio ν=0

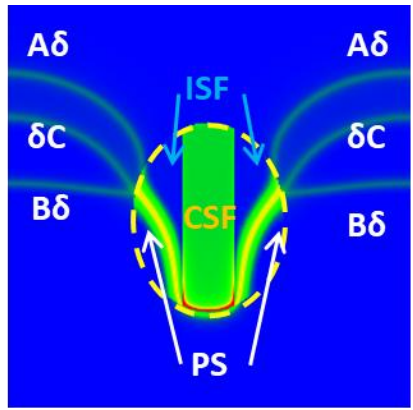

*Figure A1: Intermediate configuration of the interaction between AC+BC dislocation and the hamburger-like co-precipitate with Poisson's ratio ν=0.*

## B. Interaction between three CA dislocations and two individual γ" precipitates with no γ' between them.

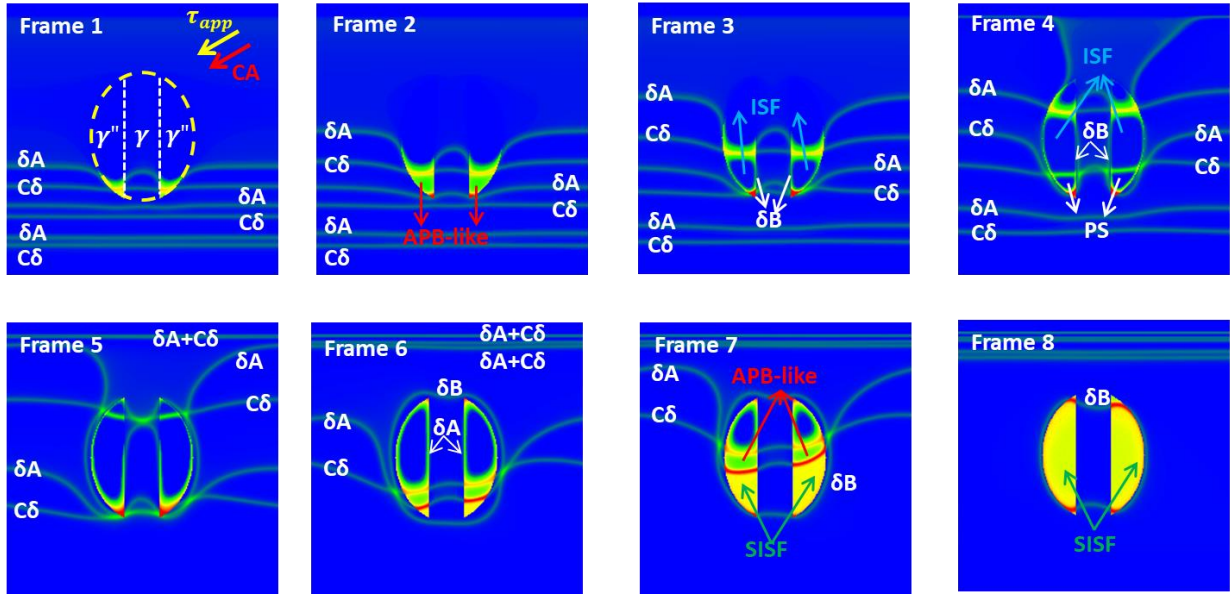

*Figure B1: Interaction between three CA dislocations and two individual γ" precipitates with no γ' between them.*

# Supplementary Material

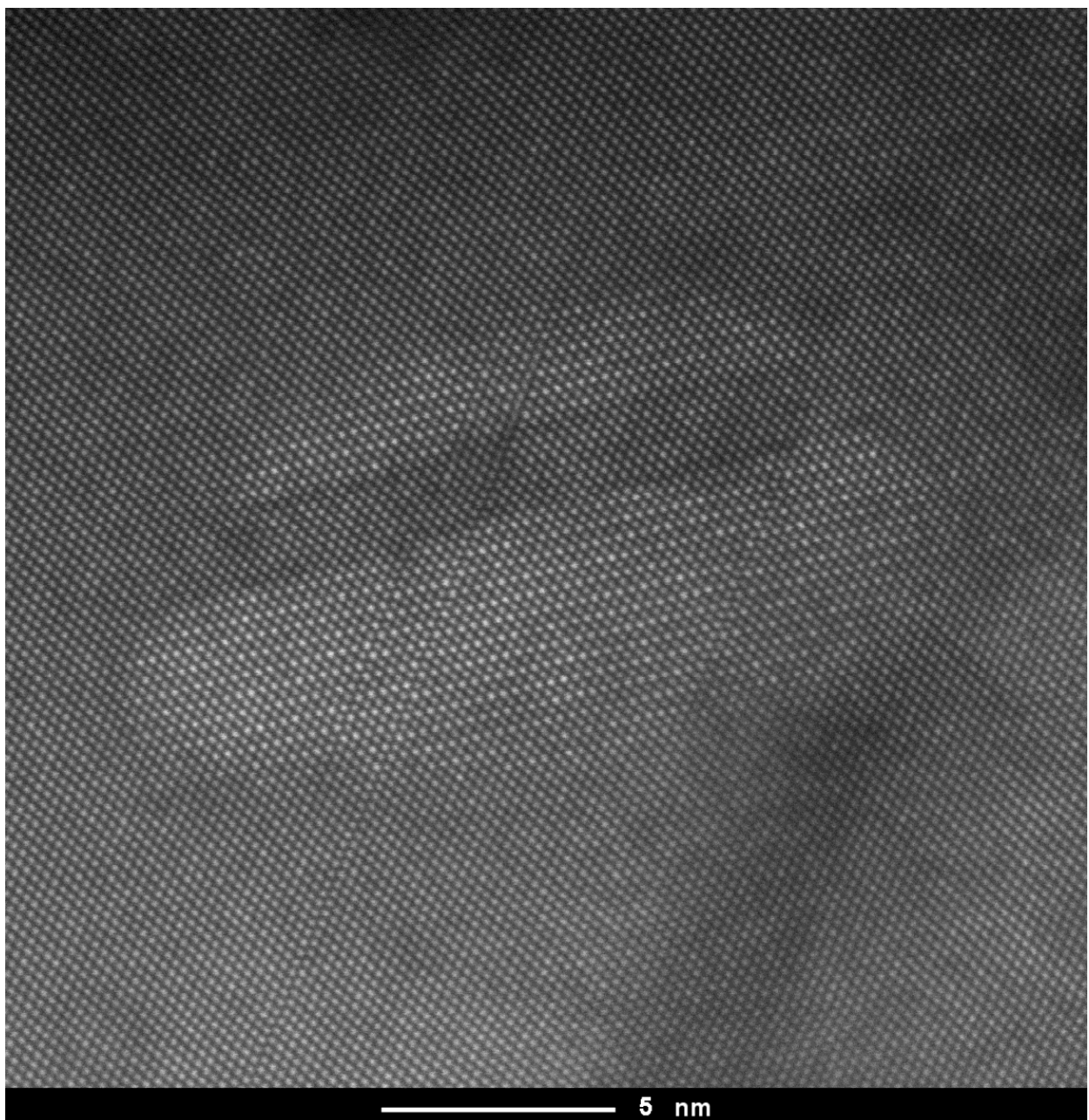

*Figure S1: Original micrograph corresponding to Figure 6.*

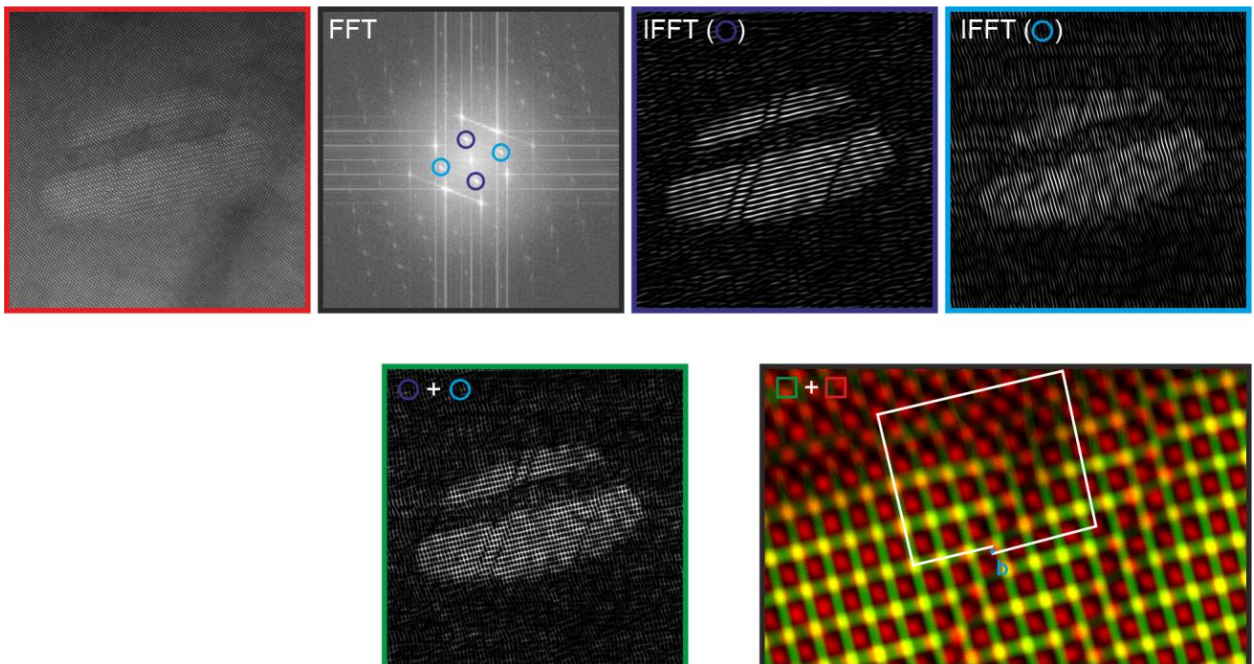

*Figure S2: Original micrograph corresponding to Figure 6.*

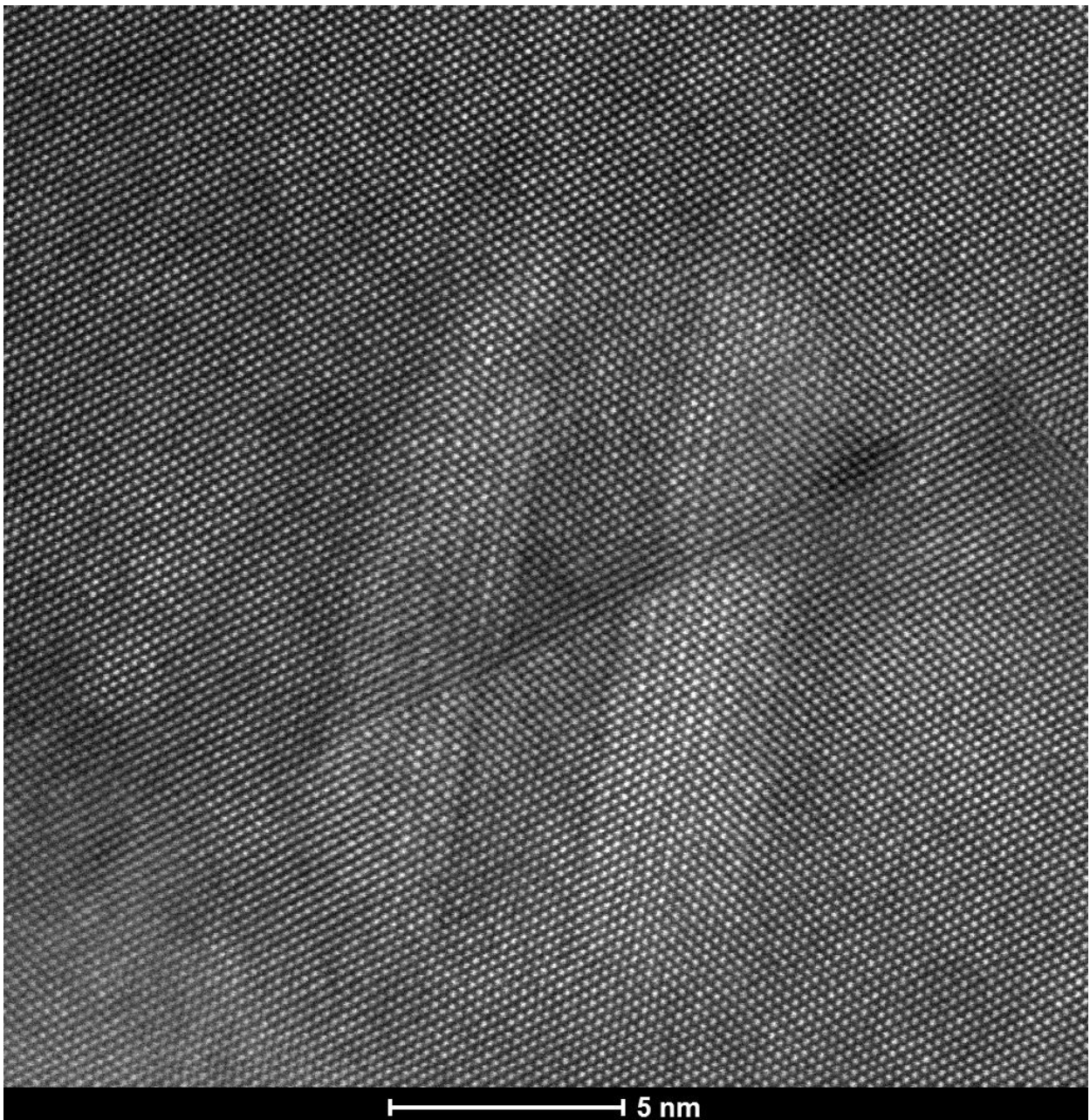

*Figure S3: Original micrograph corresponding to Figure 7.*

**Graphical Abstract**

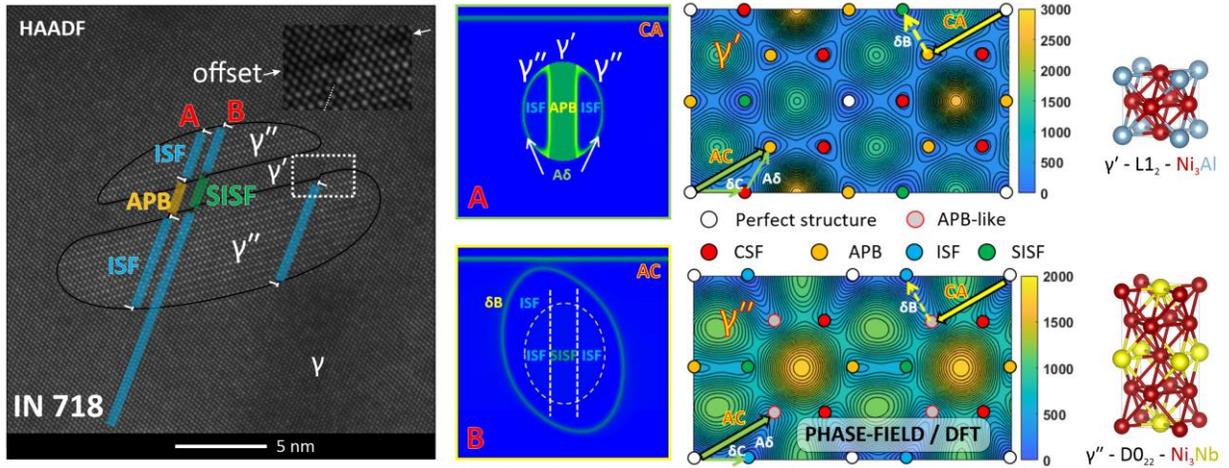